\documentclass[preprint2]{aastex61}
\usepackage{hyperref}
\usepackage{amsmath}

\shorttitle{Core Kinematics}
\shortauthors{Kong et al.}

\begin{document} 

\title{The CARMA-NRO Orion Survey: Core Emergence and Kinematics in the Orion A Cloud}

\author{Shuo Kong}
\affiliation{Department of Astronomy, Yale University, New Haven, CT 06511, USA}

\author{H\'ector G. Arce}
\affiliation{Department of Astronomy, Yale University, New Haven, CT 06511, USA}

\author{Anneila I. Sargent}
\affiliation{Cahill Center for Astronomy and Astrophysics, California Institute of Technology, 249-17, Pasadena, CA 91125, USA}

\author{Steve Mairs}
\affiliation{East Asian Observatory, 660 N. A'ohoku Place, Hilo, Hawaii, 96720, USA}

\author{Ralf S. Klessen}
\affiliation{Universit\"{a}t Heidelberg, Zentrum f\"{u}r Astronomie, Albert-Ueberle-Str. 2, 69120 Heidelberg, Germany}
\affiliation{Universit\"{a}t Heidelberg, Interdisziplin\"{a}res Zentrum f\"{u}r Wissenschaftliches Rechnen, INF 205, 69120 Heidelberg, Germany}

\author{John Bally}
\affiliation{Department of Astrophysical and Planetary Sciences, University of Colorado, Boulder, Colorado, USA}

\author{Paolo Padoan}
\affiliation{Institut de Ci\`{e}ncies del Cosmos, Universitat de Barcelona, IEEC-UB, Mart\'{i} i Franqu\`{e}s 1, E08028 Barcelona, Spain}
\affiliation{ICREA, Pg. Llu\'{i}s Companys 23, 08010 Barcelona, Spain}

\author{Rowan J. Smith}
\affiliation{Jodrell Bank Centre for Astrophysics, School of Physics and Astronomy, University of Manchester, Oxford Road, Manchester M13 9PL, UK}

\author{Mar\'ia Jos\'e Maureira}
\affiliation{Department of Astronomy, Yale University, New Haven, CT 06511, USA}

\author{John M. Carpenter}
\affiliation{Joint ALMA Observatory, Alonso de C\'ordova 3107 Vitacura, Santiago, Chile}

\author{Adam Ginsburg}
\affiliation{National Radio Astronomy Observatory,  1003 Lopezville road, Socorro, NM 87801, USA}

\author{Amelia M. Stutz}
\affiliation{Departmento de Astronom\'{i}a, Facultad de Ciencias F\'{i}sicas y Matem\'{a}ticas, Universidad de Concepci\'{o}n, Concepci\'{o}n, Chile}
\affiliation{Max-Planck-Institute for Astronomy, K\"onigstuhl 17, 69117 Heidelberg, Germany}

\author{Paul Goldsmith}
\affiliation{Jet Propulsion Laboratory, California Institute of Technology, 4800 Oak Grove Drive, Pasadena, CA 91109, USA}

\author{Stefan Meingast}
\affiliation{Department of Astrophysics, University of Vienna, T\"urkenschanzstrasse 17, 1180 Wien, Austria}

\author{Peregrine McGehee}
\affiliation{Department of Earth and Space Sciences, College of the Canyons, Santa Clarita, CA 91355, USA}

\author{\'Alvaro S\'anchez-Monge}
\affiliation{I.~Physikalisches Institut, Universit\"at zu K\"oln,
              Z\"ulpicher Str. 77, D-50937 K\"oln, Germany}

\author{S\"umeyye Suri}
\affiliation{I.~Physikalisches Institut, Universit\"at zu K\"oln,
              Z\"ulpicher Str. 77, D-50937 K\"oln, Germany}

\author{Jaime E. Pineda}
\affiliation{Max-Planck-Institut f\"ur extraterrestrische Physik, Giessenbachstrasse 1, 85748 Garching, Germany}

\author{Jo\~ao Alves}
\affiliation{Department of Astrophysics, University of Vienna, T\"urkenschanzstrasse 17, 1180 Wien, Austria}
\affiliation{Radcliffe Institute for Advanced Study, Harvard University, 10 Garden Street, Cambridge, MA 02138, USA}

\author{Jesse R. Feddersen}
\affiliation{Department of Astronomy, Yale University, New Haven, CT 06511, USA}

\author{Jens Kauffmann}
\affiliation{Haystack Observatory, Massachusetts Institute of Technology, 99 Millstone Road, Westford, MA 01886, USA}

\author{Peter Schilke}
\affiliation{I.~Physikalisches Institut, Universit\"at zu K\"oln,
              Z\"ulpicher Str. 77, D-50937 K\"oln, Germany}

\begin{abstract}

We have investigated the formation and kinematics of
sub-mm continuum cores in the Orion A molecular cloud. 
A comparison between sub-mm 
continuum and near infrared
extinction shows a continuum core 
detection threshold of $A_V\sim$ 5-10 mag.
The threshold is similar to the star formation extinction
threshold of $A_V\sim$ 7 mag proposed by recent work, 
suggesting a universal star formation extinction threshold
among clouds within 500 pc to the Sun.
A comparison between the Orion A cloud and a massive infrared
dark cloud G28.37+0.07
indicates that Orion A produces more dense gas
within the extinction range 15 mag $\la A_V \la$ 60 mag. 
Using data from the CARMA-NRO Orion Survey, 
we find that dense cores 
in the integral-shaped filament (ISF)
show sub-sonic core-to-envelope velocity dispersion that is
significantly less than the local envelope line dispersion,
similar to what has been found in nearby clouds.
Dynamical analysis indicates that the cores are
bound to the ISF. An oscillatory core-to-envelope 
motion is detected along the ISF. Its origin is to 
be further explored.

\end{abstract}


\section{Introduction}

It is generally accepted that stars form in gravitationally bound 
dense cores within molecular clouds. 
Understanding the formation and evolution of cores is therefore
crucial to the study of star formation (SF). For example,  
observations \citep[e.g.,][]{2010ApJ...724..687L,2010ApJ...723.1019H} 
and theory \citep[e.g.,][]{1989ApJ...345..782M}
imply that the onset of the star formation process
occurs at the transition between the cloud
photo-dissociation region (PDR) and self-shielded regions,  
where the cloud extinction is at least $A_V\sim$ 7 mag.
It seems reasonable to expect a similar 
extinction threshold for core formation (CF).
Other outstanding problems in SF concern the efficiency 
of the overall process \citep{2012ARA&A..50..531K}, and
the effect of core kinematics on the dynamics of emerging clusters.

At a distance of 400 pc, 
the Orion A cloud provides an opportunity to investigate
 the early formation of cores in details.
In particular, Orion A enables a study of the effects of
feedback from new young massive stars on their birth environment.
Here, we combine newly acquired $^{12}$CO(1-0), $^{13}$CO(1-0), 
and C$^{18}$O(1-0) molecular line data cubes of extended regions
on the Orion A cloud from the CARMA-NRO Orion Survey
\citep[][hereafter K18]{2018ApJS..236...25K}
with a variety of complementary surveys at other wavelengths 
to address these issues. The new high dynamic range images
recover spatial scales from  $\sim$8\arcsec\ (0.015 pc)
to $\sim$2.5\arcdeg\ (18 pc). 

As noted above, it would be useful to compare the extinction
thresholds between SF and CF.
Millimeter continuum observations of the Ophiuchus 
and Perseus molecular clouds show evidence
of a CF threshold at $A_V\sim$ 5-9 mag
\citep{2004ApJ...611L..45J,2006ApJ...644..326Y,
2006ApJ...638..293E,2006ApJ...646.1009K}. However,  
see \citet{2014MNRAS.444.2396C} for a critical discussion. 
The Orion A cloud can be a unique test of the extinction 
threshold of CF because, unlike the other nearby clouds mentioned above, it has strong 
feedback from  young massive stars. It is important to 
show if the CF threshold varies in drastically different
environments. 

Above the extinction threshold, the SF efficiency is of 
order 10\% \citep{2010ApJ...724..687L,2010ApJ...723.1019H}.
This means that SF is still limited even when the cloud
extinction is above the so-called threshold. This has 
long been known as a challenge in SF \citep{2012ARA&A..50..531K}.
In our context, it is interesting to see whether the 
low efficiency is already evident in the formation 
of cores. Unlike the observations of
SF efficiency where one counts young stellar objects (YSOs),
and may miss many of those which have left their 
birthplace, the star-forming cores
are still embedded in the host cloud and
maintain the primordial information of the core/star formation.
This allows  a more thorough examination of the
dependence of the SF efficiency on  natal environment.

Core kinematics bring another type of insights to our
understanding of CF. In particular, core-to-core kinematics,
potentially shaped by the host cloud, may have
a notable impact on the dynamics of the forthcoming star cluster.
With our newly acquired molecular line data from 
the CARMA-NRO Orion Survey (K18),
we are able to investigate the core kinematics in great details.
Throughout the paper, we follow K18  and use a distance to Orion A
of 400 pc \citep[see][for the latest discussions]{2018AJ....156...84K,
2018A&A...616A...1G}. In the following, we introduce our data collection
in \S\ref{sec:data}, and our results and analysis in \S\ref{sec:results}.
Finally, we present the discussion and conclusions in 
\S\ref{sec:discussions} and \S\ref{sec:conclusions}, respectively.

\section{Data Collection}\label{sec:data}

\subsection{CARMA-NRO Orion Data}\label{subsec:carma}

In this paper, we use the molecular line data from the
CARMA-NRO Orion Survey (see K18 for more details). The survey
produced cubes for $^{12}$CO(1-0), $^{13}$CO(1-0), and C$^{18}$O(1-0).
We combined single-dish data from the Nobeyama 
45m telescope (NRO45) and interferometer data from the Combined Array 
for Research in Millimeter Astronomy (CARMA), producing spectral maps that recover spatial scales from $\sim$8\arcsec\ (0.015 pc)
to $\sim$2.5\arcdeg\ (18 pc). The velocity resolution is 
0.25 km s$^{-1}$ for $^{12}$CO(1-0); 0.11 km s$^{-1}$ for 
$^{13}$CO(1-0) and C$^{18}$O(1-0). 

\subsection{Near-infrared Extinction Map}\label{subsec:vista}

\begin{figure*}[htbp]
\epsscale{1.17}
\plotone{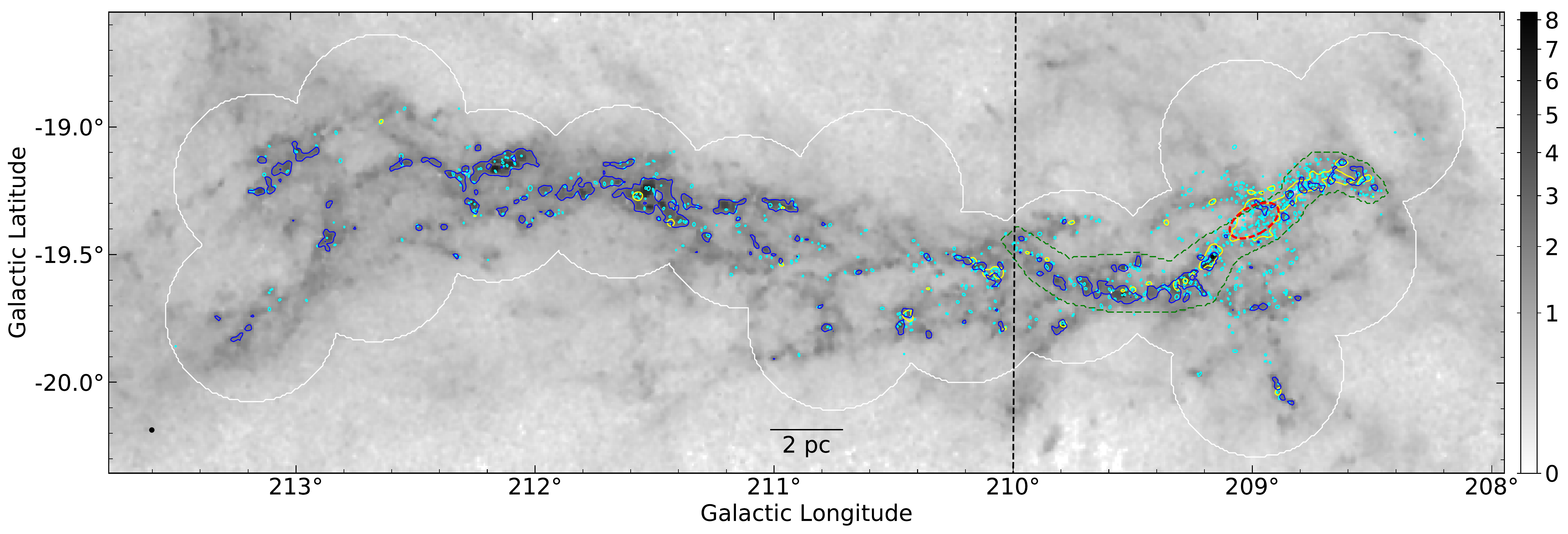}
\caption{
{\it Greyscale:} Meingast18 K-band extinction map.
{\it Yellow contours:} Lane16 
850 $\mu$m continuum image smoothed to 1\arcmin\ resolution.
The contour level is $\Sigma_{\rm mm}$ = 0.044 g cm$^{-2}$ 
(see \S\ref{subsec:dpfirdc}). {\it Blue contours:} 
Meingast18 K-band extinction equivalent to $A_V = 20$ mag.
{\it Cyan contours:} Lane16 Getsources cores.
{\it White contour:} Lane16 map coverage.
{\it Green polygon:} The definition of ISF in 
Figure \ref{fig:filregions} (see \S\ref{subsec:core}).
The red dashed ellipse shows the region
that is masked for the detection probability study.
The 1\arcmin\ beam size is shown at the lower-left corner. 
The vertical dashed line splits the cloud into two regions
with different level of feedback from massive stars. 
See \S\ref{subsec:dpfirdc}.
\label{fig:LaneMeingast}}
\end{figure*}

We use the near infrared (NIR) extinction map of Orion A
recently published by \citet[][hereafter Meingast18]{2018A&A...614A..65M}
to probe the cloud's dust column density distribution
at a resolution of 1\arcmin (or 2400 au),
and a pixel scale of 30\arcsec\ (see Figure \ref{fig:LaneMeingast}).
For our study, we mask the vicinity of the
Orion KL region to avoid underestimated extinctions.
We convert the map from $A_K$ to 
$A_V$ by multiplying by a factor of 8.93 
\citep{1985ApJ...288..618R}.
In the following, the extinction map is masked to
have the same coverage as the continuum image.

\subsection{Sub-mm Continuum Image and Core Catalog}\label{subsec:jcmt}

We obtain the 850 $\mu$m continuum image from 
\citet[][hereafter Lane16]{Lane2016}
for our study (publicly available at 
\url{https://doi.org/10.11570/16.0008}). 
The image was acquired as part of the JCMT Gould Belt Survey
\citep{2007PASP..119..855W,2015MNRAS.449.1769S,2016MNRAS.461.4022M}. 
The 850 $\mu$m map has a resolution of 14.6\arcsec\ ($\sim$0.03 pc).
Below we investigate the relation between continuum detection
and cloud column density traced by the Meingast18 extinction map.
First, we compare the continuum core catalog
defined by Lane16 (the Getsources catalog) with
the Meingast18 extinction map\footnote{One 
caveat is that the cores were defined with 
14.6\arcsec\ resolution while the extinction 
map resolution is  1\arcmin.}.
Second, we carry out a pixel-by-pixel
comparison between the Lane16 sub-mm continuum image
and the Meingast18 extinction map without defining ``cores''.
Before the second comparison, we
convolve the dust continuum image with a Gaussian kernel to 
have a final spatial resolution of
1\arcmin\ and regrid it to the Meingast18
extinction map. The smoothed image has an rms 
noise of $\sigma_{850}$ = 10 mJy per 1\arcmin\ beam.

Lane16 used an automask data reduction technique for the 
continuum image. This is a reduction strategy wherein 
astronomical flux (as opposed to atmospheric flux or noise) is 
identified automatically in the map-making procedure based on 
pixel SNR and the results of iteratively calculated noise models. 
For further details on the data reduction, 
see \cite{2013MNRAS.430.2545C},
\cite{2015MNRAS.454.2557M}, and Lane16.
As a result, sources with sizes between 2.5\arcmin\ and 
7.5\arcmin\ (0.3 pc and 0.9 pc at a distance of 400 pc)
do not have robust flux measurements. Sources smaller than
2.5\arcmin\ likely have robust flux measurements.
Based on the work in \citet{2015MNRAS.454.2557M},
Lane16 indicated that strong, compact
sources are well recovered, while faint, diffuse 
sources suffer more from flux loss. 
Since the goal is to target cores that are more centrally
concentrated, i.e., more likely to develop protostars,
we consider that the sub-mm continuum image 
satisfactorily represents the emerging cores in Orion A.

\subsection{GAS Ammonia Core Catalog}\label{subsec:gbt}

We have obtained the NH$_3$ data from the GAS survey
\citep[][hereafter Kirk17]{2017ApJ...843...63F,2017ApJ...846..144K}, 
which is publicly available at
\url{https://dataverse.harvard.edu/dataverse/GAS_DR1}.
Kirk17 selected a sample of cores from the larger
sample of Lane16 based on the detection of NH$_3$ 
\citep[observed with the Green Bank Telescope 
in the GAS survey, see][]{2017ApJ...843...63F} 
and studied their physical properties. 
In total, 237 continuum cores, most of them in the ISF,
were included in their study.
Kirk17 performed hyperfine line
fitting to the NH$_3$ inversion lines and estimated 
the NH$_3$ core velocity $v_{\rm NH_3}$ and 
the kinetic temperature based on the ammonia emission. 
The GAS survey resolution is 32\arcsec, 
larger than some of the sub-mm cores in Lane16. Kirk17 (see 
their \S3.2) have argued that the NH$_3$ traces the 
dense cores reasonably well. In this paper, 
we adopt the assumption that the dense core line-of-sight
velocity is traced by the Kirk17 NH$_3$ line fitting result.



\section{Results and Analyses}\label{sec:results}

\subsection{Sub-mm continuum detection}\label{subsec:dpf}

\begin{figure}[htbp]
\epsscale{1.}
\plotone{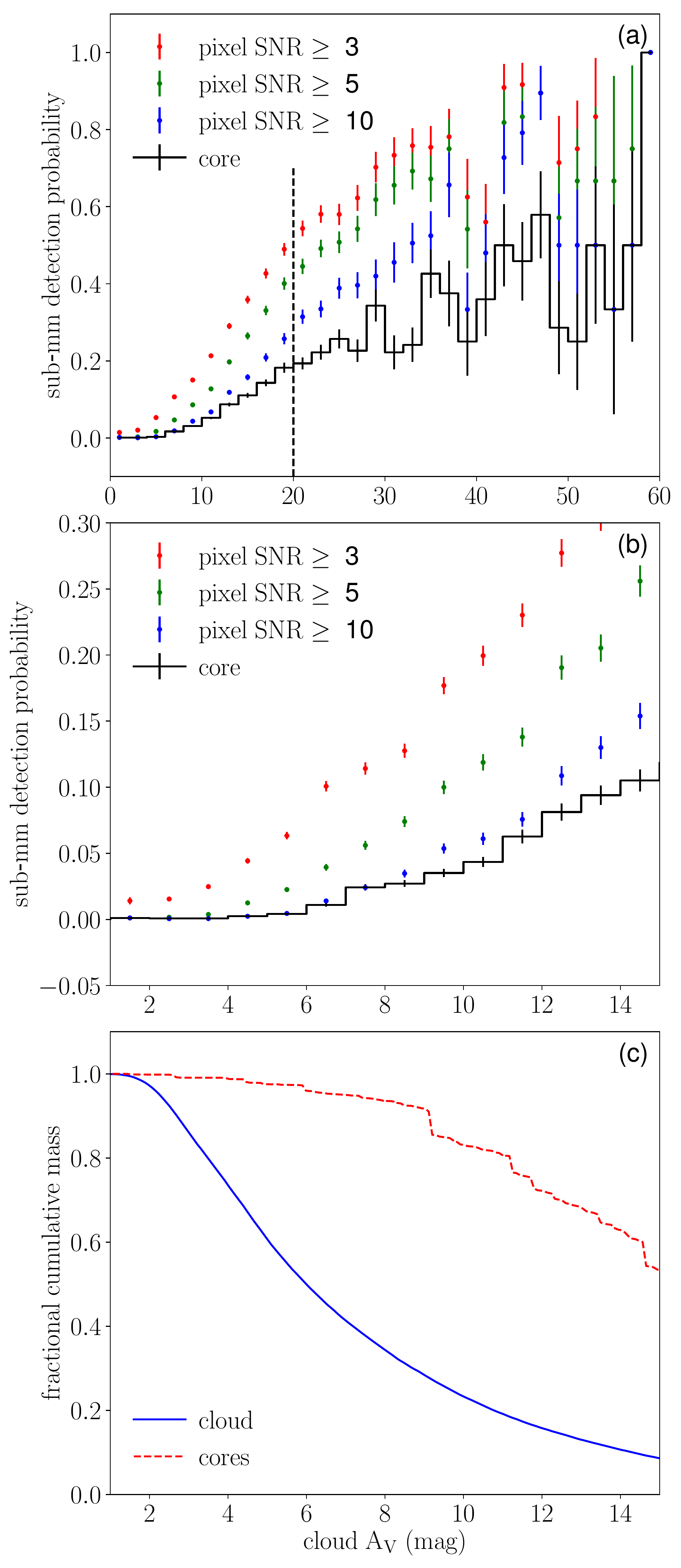}
\caption{
Continuum detection probability for Orion A. 
{\bf (a):} The black histogram shows the core 
detection probability as a function of $A_V$.
The color points are pixel detection probability functions.
The continuum detection thresholds are indicated
in the top-left legends. The vertical dashed line
indicates the potential break. The bin size is
2 mag in $A_V$.
{\bf (b):} Same as panel (a), 
but zoomed in for $1<A_V<15$ mag.
The bin size is 1 mag, compared to 2 mag for panel (a).
{\bf (c):} Cumulative distribution of extinction
for the cloud (blue) and the cores (red).
For each $A_V$, the distribution is the fraction of 
pixels/cores with extinction greater than $A_V$.
\label{fig:dpf}}
\end{figure}

Figure \ref{fig:dpf}(a) shows the continuum detection
probability in Orion A. First, we investigate how
likely the Lane16 cores are detected at a given $A_V$.
We make $A_V$ bins every 2 mag.
In each bin, the detection probability of the cores 
$P_{\rm c,850~\mu m}$
is defined as the number of pixels 
that contain cores ($N_c$) divided by the
number of total pixels in the extinction
bin $N_{\rm tot}$, i.e., 
\begin{equation}\label{eq:pc850}
P_{\rm c,850~\mu m}~=~N_c/N_{\rm tot}.
\end{equation}
The error is estimated as 
\begin{equation}\label{eq:epc850}
\sigma_P=[P_{\rm c,850~\mu m}(1-P_{\rm c,850~\mu m})/N_{\rm tot}]^{0.5}.
\end{equation}
This estimation gives zero error when $P_{\rm c,850~\mu m}$
is 0 or 1 \citep[see discussions in][]{2018ApJ...855L..25K}.
Figure \ref{fig:dpf}(b) zooms in to 1 mag $<A_V<$ 15 mag,
with a bin size of 1 mag.

Figure \ref{fig:dpf}(a) shows that the detection 
probability of the Lane16 continuum cores remains 0 for
$A_V~\leq~5$ mag. At $A_V~>~5$ mag, the probability
increases monotonically until $A_V~\sim~20$ mag,
after which it has a shallower slope. 
In panel (b), the core detection (histogram)
shows an increase after $A_V~\sim~7$ mag.
Following \citet[][figure 7]{2016MNRAS.461.4022M}, 
we make a cumulative distribution function
(CDF) for the cloud and core extinction, shown in
the panel (c). The core CDF remains
approximately flat until a turnover at $A_V\sim9$ mag.
Based on these, we assign a value to the core
detection threshold in the range $A_V \sim 5-10$ mag.
This is consistent with findings in
Ophiuchus \citep[9 mag,][]{2006ApJ...644..326Y}
and Perseus \citep[5 mag,][]{2006ApJ...638..293E}.
As a massive star-forming cloud, Orion A 
does not show a significantly different
extinction threshold of core formation compared
to other lower-mass clouds.
Moreover, these results are also consistent with 
\citet{2008ApJ...680..428G,2010ApJ...724..687L,2010ApJ...723.1019H}
who investigated the SF extinction threshold using YSOs.
The similarity between the core formation threshold
and the SF threshold is not unexpected since stars 
form in dense cores.
These comparisons potentially suggest a 
universal SF extinction threshold of $A_V=5-10$ mag
in the clouds within 500 pc of the Sun.

It is also important to study 
continuum detections that are not identified as
cores. They may be the precursors of cores as
they become centrally peaked during fragmentation. 
We show in Figure \ref{fig:dpf}(a)(b) the pixel-based
850 $\mu$m detection probability function (DPF).
The pixel detection probability is the ratio between
the number of extinction pixels that contain 850 $\mu$m 
dust emission detection and the number 
of total pixels within each extinction bin.
Three continuum detection thresholds are applied, 
namely, SNR$\geq$3 (red); SNR$\geq$5 (green);
SNR$\geq$10 (blue). The ``detected'' pixels defined
this way may include noisy spikes in the image.
Using a higher threshold can reduce
a false detection at the expense of missing real emission.   

As expected, the pixel-based continuum detection
probability is larger than the core-based probability,
since the cores are defined with a subset of detected
pixels (Lane16). In fact, a notable number
of the detected continuum pixels are not included in
the cores defined by Lane16 {\it Getsources} catalog.
As described in their appendix A.1.1, Lane16 set a
threshold SNR = 7 for core detection. Further, they
smoothed the 850 $\mu$m image and removed sources 
that did not appear significant, along with those
rejected by the {\it Getsources} method. As a result,
the number of ``reliable'' cores was reduced from 1178
to 919 (Lane16 final catalog). 
Figure 2 of Lane16 shows that only a fraction of the
emission was grouped into cores. 

As shown in Figure \ref{fig:dpf}(a)(b), 
the DPF rises from 0
to close to 1 from $A_V=0$ mag to $A_V\sim 50$ mag.
At a given extinction, only a fraction of the pixels
are detected in continuum emission, caused by a combination
of low temperature, spatial filtering effect,
and/or opacity variation. At sub-mm and mm wavelengths,
the flux density is approximately
proportional to the column density and the temperature.
The line-of-sight column temperature may have a major
impact on the detection above a given threshold. 
The cores are likely heated
by contraction and/or an embedded protostar
\citep[see also discussions in][]{2018ApJ...855L..25K}.
Spatial filtering of large spatial structures
 helps in picking out centrally peaked dense cores.
Meanwhile, the extinction traces the total column density.
Therefore, we argue that the continuum detection
pinpoints the forming dense cores embedded in the extinction column density.

The DPF thus shows that only a fraction of the cloud is 
forming stars at a given time, consistent with the
theoretical expectation in, e.g., \citet{2005ApJ...630..250K}.
An interesting feature in Figure \ref{fig:dpf}(a)
is the apparent break at $A_V\sim 20$ mag
(vertical dashed line). For the DPF with SNR$\geq$10,
a linear regression for $10 \la A_V \la 20$ mag gives
a slope of 0.024($\pm$0.001), while the slope for
$20 \la A_V \la 30$ mag is 0.012($\pm$0.001).
Interestingly, $A_V = 20$ mag is also where the core detection 
slope becomes shallower. For reference,
Figure \ref{fig:LaneMeingast} shows the regions with
$A_V = 20$ mag with blue contours. It remains to be
seen if the 20 mag break is detected in other clouds.

The total mass of the Lane16 cores is $\sim$ 1400 M$_\odot$.
We follow Lane16 equation (1) for the mass calculation.
This was also adopted by Kirk17.
Note this number has been scaled down by a factor of 0.8
compared to the Lane16 and Kirk17 core mass because
we use 400 pc as the cloud distance while they used 
450 pc. The total cloud mass converted from the 
Meingast18 extinction map is $\sim$ 36,000 M$_\odot$
within the Lane16 mapped area.
The Lane16 core mass fraction is therefore $\sim$ 4.0\% 
of the mass estimated from the extinction map.
If we include all Lane16 sub-mm continuum flux above a
SNR of 5, the mass fraction increases to $\sim$ 
14\%\footnote{The 15 K assumption for the dust temperature
around Orion KL is probably an underestimate, which results in an
overestimate of the mass fraction. Excluding the KL
region, we find a mass fraction of $\sim$ 4.0\%.}.

\subsection{Comparison with IRDC G28.37+0.07}\label{subsec:dpfirdc}

\begin{figure}[htbp]
\epsscale{1.2}
\plotone{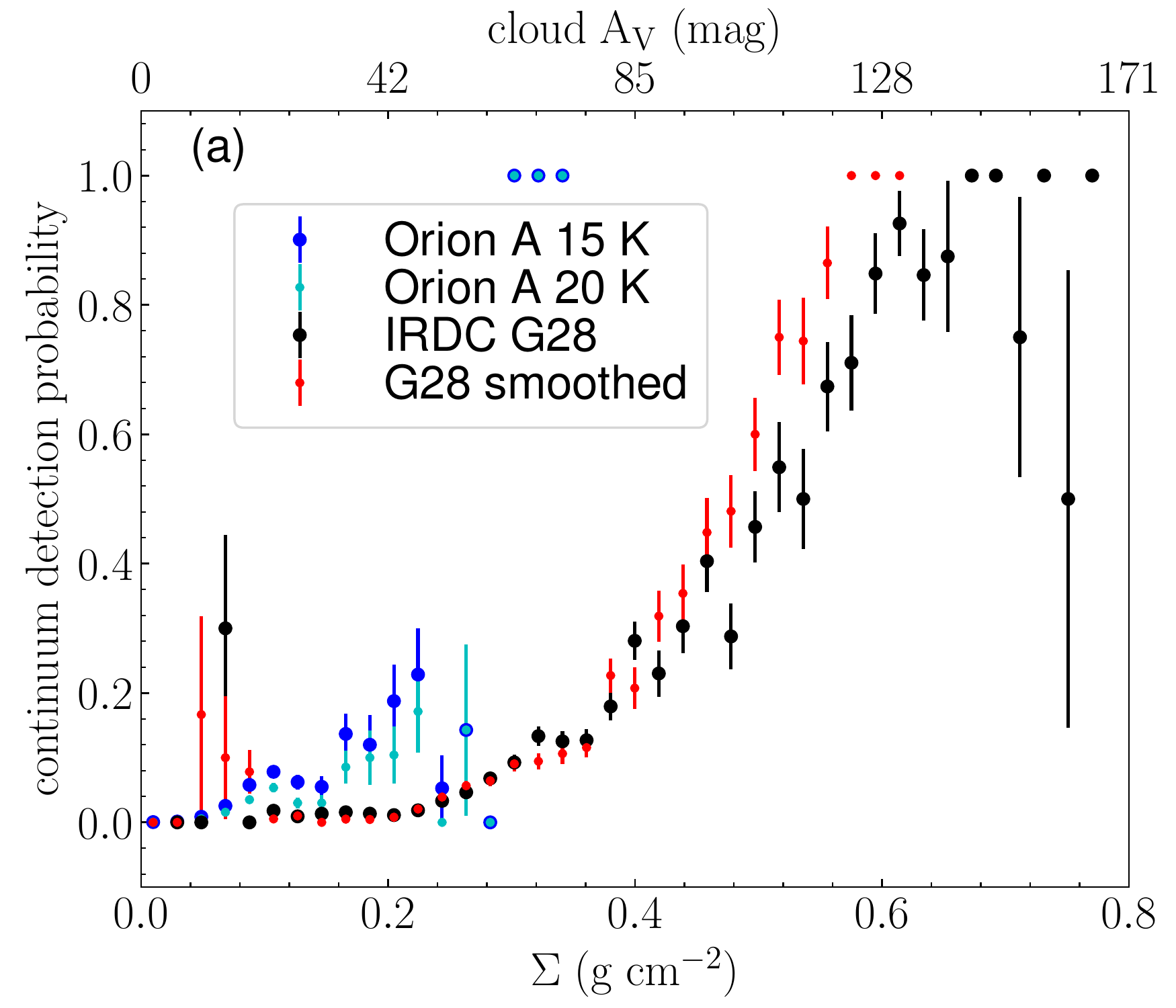}\\
\plotone{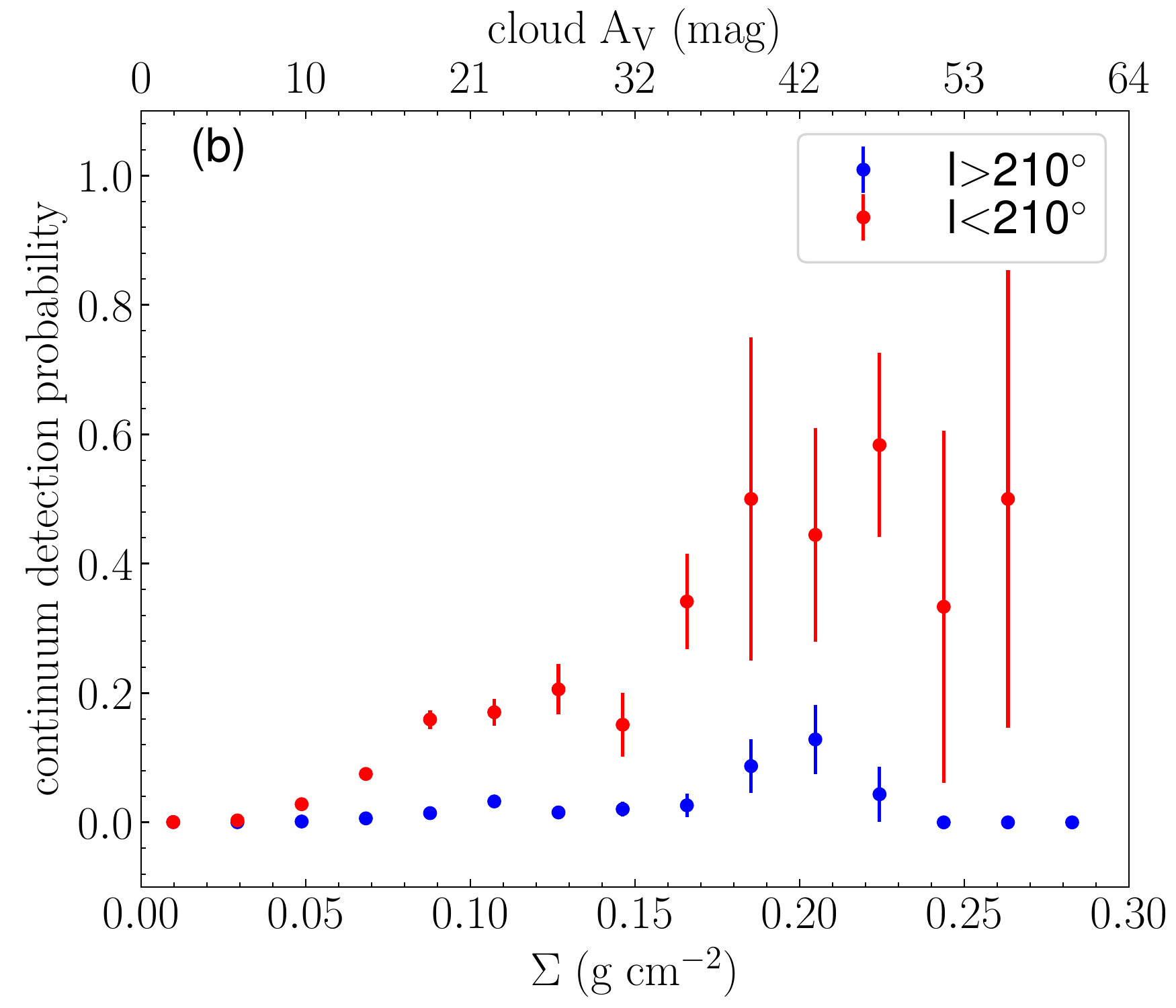}
\caption{
\textbf{(a): } 
Continuum pixel detection probability for
Orion A (blue) and IRDC G28 (black). 
The red points are from smoothed G28 data
that matches the physical resolution of
the Orion A data. The cyan points are
the same as the blue ones except that we
use 20 K to calculate the detection flux
threshold. See \S\ref{subsec:dpfirdc}.
\textbf{(b): }
Continuum pixel detection probability for
two sub-regions in Orion A (divided at 
l=210\arcdeg, indicated by the vertical dashed 
line in Figure \ref{fig:LaneMeingast}). 
\label{fig:dpfcompare}}
\end{figure}

To investigate the dense gas emergence in different
environments, we compare the Orion A DPF with the DPF
from the infrared dark cloud (IRDC) G28.37+0.07 (hereafter G28).
\citet{2018ApJ...855L..25K} studied the DPF in IRDC G28
with ALMA 1.3 mm continuum data. Although the DPF in Orion A
is derived with 850 $\mu$m continuum, the comparison is still
meaningful once we apply the same physical threshold.
For comparison, the physical resolution is 0.05 pc
in the IRDC G28 study \citep{2018ApJ...855L..25K}.
The maximum detectable scale was 0.5 pc.
In IRDC G28, assuming a temperature of 15 K and volume
density of 10$^5$ cm$^{-3}$, the Jeans length 
is $\sim$ 0.1 pc. 
We also smooth the G28 data to a resolution of 0.12 pc, 
in order to match the linear resolution of the 
IRDC data with that of the Orion A study. 
The fiducial 1.3 mm continuum detection threshold in mass surface 
density was set to $\Sigma_{\rm mm}$ = 0.044 g cm$^{-2}$ in 
\citet{2018ApJ...855L..25K}. To match this number
in Orion A, we compute the mass surface density following
equation (1) in Lane16 and adopting the same assumptions
of dust temperature and opacity. The threshold is
shown as the yellow contours in Figure \ref{fig:LaneMeingast}.

Figure \ref{fig:dpfcompare}(a) shows the comparison of
DPFs between Orion A and IRDC G28.
We use the same bin size of 0.02 g cm$^{-2}$
for both clouds, following \citet{2018ApJ...855L..25K}.
Note that this number is converted from extinction 
(hereafter denoted $\Sigma_{\rm ex}$),
which is different from the sub-mm continuum detection
threshold (also in g cm$^{-2}$, but from emission).
The highest extinction traced by the Meingast18 map
is about 0.3 g cm$^{-2}$. Recently, 
\citet{2018MNRAS.473.4890S} used Herschel far infrared
and APEX 870 $\mu$m data to derive a column
density map in Orion A. The highest surface density
reaches $\sim$ 3 g cm$^{-2}$
in the Orion KL region. The near infrared extinction 
does not work well in this region. However, the number of 
pixels in this region only accounts for $\sim$0.1\% of the
Lane16 coverage. We ignore these pixels
by removing them from the comparison (red dashed ellipse
in Figure \ref{fig:LaneMeingast}).

As shown in Figure \ref{fig:dpfcompare}(a), 
the Orion A cloud shows more continuum detection of dense gas with 
$\Sigma_{\rm mm}\geq$ 0.044 g cm$^{-2}$ than IRDC G28. 
The difference appears within 0.08 $\la\Sigma_{\rm ex}\la$ 0.28 g cm$^{-2}$. 
For this we  followed Lane16 and Kirk17 and assumes  dust temperature of 15 K.
\citet{2018ApJ...855L..25K} assumed 20 K for their dust 
continuum emission because they argued that the cores were
mostly protostellar. For consistency, we also calculate a 
new DPF assuming a dust temperature of 20 K in Orion A
(cyan dots in Figure \ref{fig:dpfcompare}(a)).
A higher dust temperature requires a higher continuum flux
to reach the same mass surface density threshold of
0.044 g cm$^{-2}$, therefore the cyan detection 
probability becomes lower. Nonetheless, the Orion A
DPF is still higher than the G28 DPF within 
0.08 $\la\Sigma_{\rm ex}\la$ 0.28 g cm$^{-2}$. 
A two-sample KS test gives a
KS statistic of 0.8 with a p-value of 0.0012,
meaning the null hypothesis that the two samples 
are drawn from the same distribution can be rejected
at a significance level below 1\% (confidence over 99\%).

\begin{figure*}[htbp]
\epsscale{1.1}
\plotone{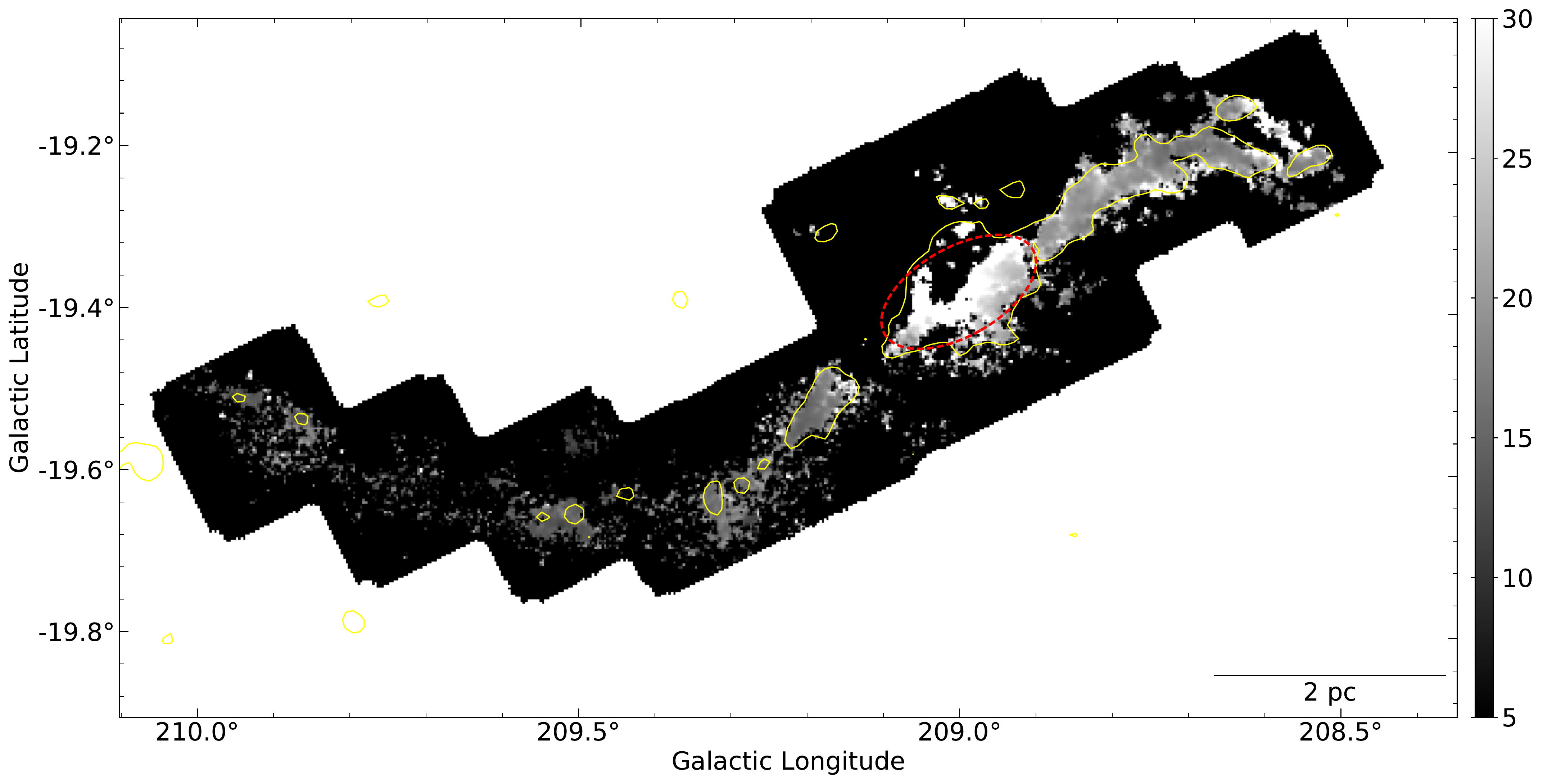}
\caption{
Kirk17 NH$_3$ gas kinetic temperature map.
The color bar ranges linearly from 5 K to 30 K.
The yellow contours and the red ellipse are the 
same as in Figure \ref{fig:LaneMeingast}.
\label{fig:tkinmap}}
\end{figure*}

The main source of uncertainty is likely the temperature
assumption. For a given mass, a higher dust temperature 
indicates a higher continuum flux. Therefore, we need 
to set a higher flux threshold for the same mass threshold.
This gives rise to lower detection
probability, which is  illustrated by the blue
and cyan points in Figure \ref{fig:dpfcompare}(a).
Lane16 pointed out that the high dust temperature 
from Herschel results \citep{2014A&A...566A..45L}
likely trace the dust in lower density regions along the 
line-of-sight and thus underestimate the core mass. 
In Figure \ref{fig:tkinmap},
we show the Kirk17 NH$_3$ kinetic temperature map.
Again, the yellow contours show the regions with a
continuum surface density equal to or greater than
0.044 g cm$^{-2}$.
The regions enclosed by the yellow contours
in Figure \ref{fig:dpfcompare}
are dominated by temperatures below 20 K.

The effect of spatial filtering may be important
when comparing the DPF in both clouds. 
Both the ALMA data and the JCMT
data filter out emission from structures at large spatial scales.
Therefore, the continuum emission in both maps detect  
relatively compact structures, that presumably trace
the dense gas intimately involved in the core formation process.  
In the JCMT 850 $\mu$m image, structures larger than 2.5\arcmin\
(0.3 pc) are not robustly recovered. In the ALMA 1.3 mm image, 
the maximum sensible scale is 12\arcsec\ \citep[0.3 pc, 
which is 0.6 times the maximum scale that corresponds to
the shortest baseline, see][]{2018ApJ...855L..25K}.
Therefore, we argue that both DPFs are tracing similar dense structures,
and Orion A is more capable of forming dense gas/cores than the IRDC G28.

One significant difference between G28 and Orion A is that 
the former is likely forming the very first generation of 
stars in the mapped region. In particular, the area studied
by \citet{2018ApJ...855L..25K} is mostly dark up to 70 $\mu$m. 
As opposed to Orion A, there is no feedback 
(radiation, spherical wind/bubble) from 
massive stars in IRDC G28, although massive stars 
may eventually form there 
\citep{2015ApJ...804..141Z,2018ApJ...867...94K}.

To test if the high DPF in Orion A is related to feedback,
we split the Orion A map in two,
indicated by the vertical dashed line along l=210\arcdeg\ in
Figure \ref{fig:LaneMeingast}. The region on the east
side (l$>$210\arcdeg) is less influenced by feedback
from the Trapezium cluster compared to the region on 
the west side \citep[l$<$210\arcdeg,][]{2008hsf1.book..459B}. 
As shown in Figure \ref{fig:dpfcompare}(b),
the high sub-mm continuum detection probability 
in Orion A is dominated by the l$<$210\arcdeg
sub-region, suggesting that  strong feedback
may indeed result in higher values in the DPF
for the same cloud extinction (see \S\ref{subsec:cf}). 

\subsection{Core-to-envelope velocity}\label{subsec:kin}

We investigate the relative motion between the 
dense cores traced by NH$_3$ (\S\ref{subsec:gbt}) and
the ambient cloud traced by CO isotopologues
(\S\ref{subsec:carma}). Similar studies include
\citet{2004ApJ...614..194W} and \citet{2010ApJ...723..457K}.
We first convolve the CARMA-NRO cubes to 32\arcsec\ 
angular resolution to match the GAS data.
Then, for  each core we extract a spectrum averaged over the 
32\arcsec\ circular area centered on the Kirk17 core position.
We fit a Gaussian to the spectra in order to derive
the bulk gas velocity component (1D) 
along the line of sight, 
with the optically thin assumption.
In fact, we have shown in K18 that more 
than 99\% of the $^{13}$CO pixels have 
$\tau<$1 (see their \S 5.2). The maximum core optical 
depth for the $^{13}$CO(1-0) line is 1.12
and 98\% of the cores have optical depth $<$ 1.

\begin{figure}[htbp]
\epsscale{1.1}
\plotone{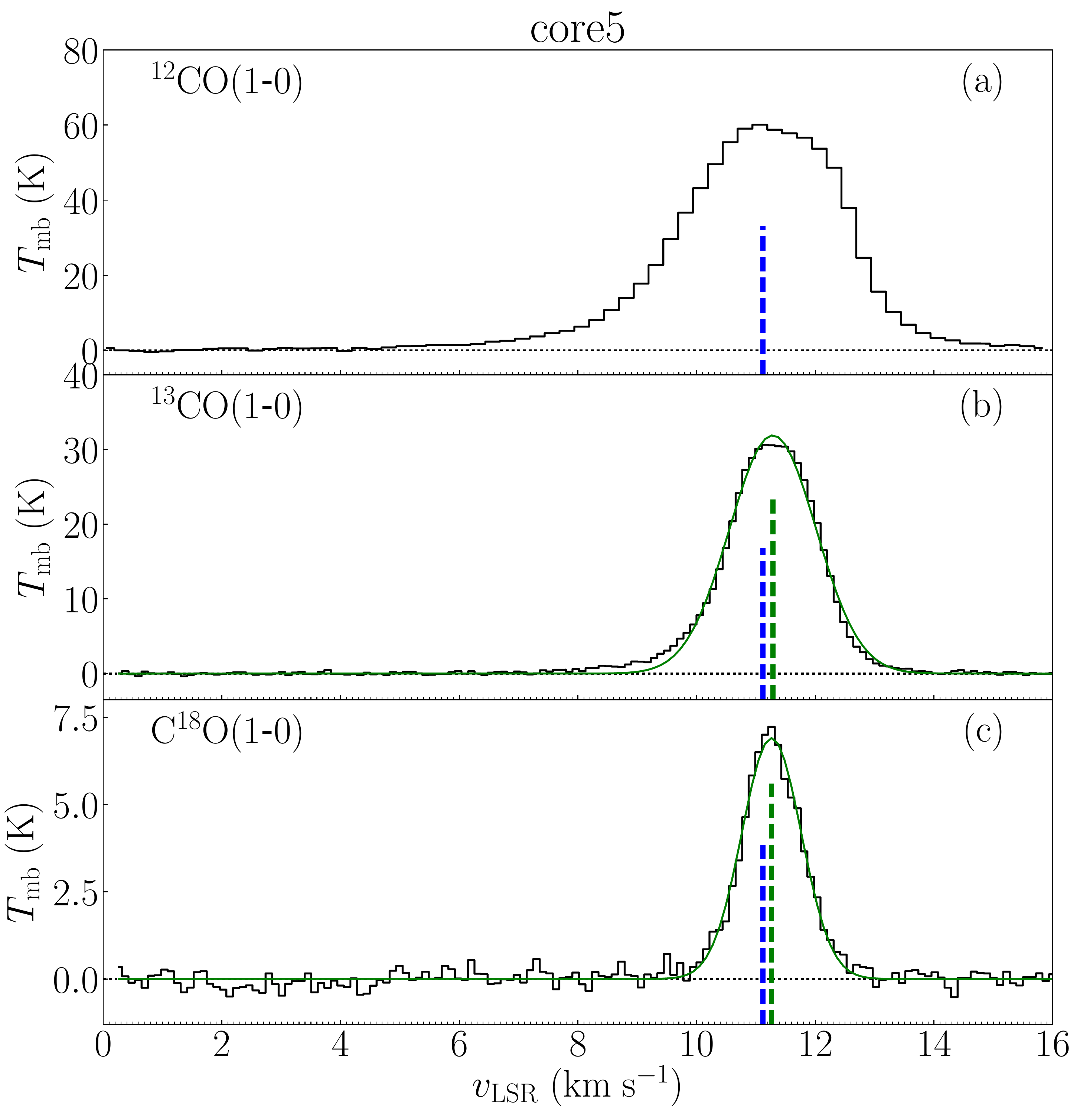}
\caption{
$^{12}$CO(1-0) ({\it top}), $^{13}$CO(1-0) ({\it middle}), 
and C$^{18}$O(1-0) ({\it bottom}) spectra for core 5 in Kirk17.
The baselines are shown as the black horizontal dashed lines.
The blue vertical dashed lines indicate $v_{\rm NH_3}$. 
The green curves show the Gaussian fit to the spectra
(only for $^{13}$CO and C$^{18}$O),
with the green vertical dashed lines showing the 
centroid  velocities $v_{\rm gauss}$.
\label{fig:core5}}
\end{figure}

\begin{figure}[htbp]
\epsscale{1.1}
\plotone{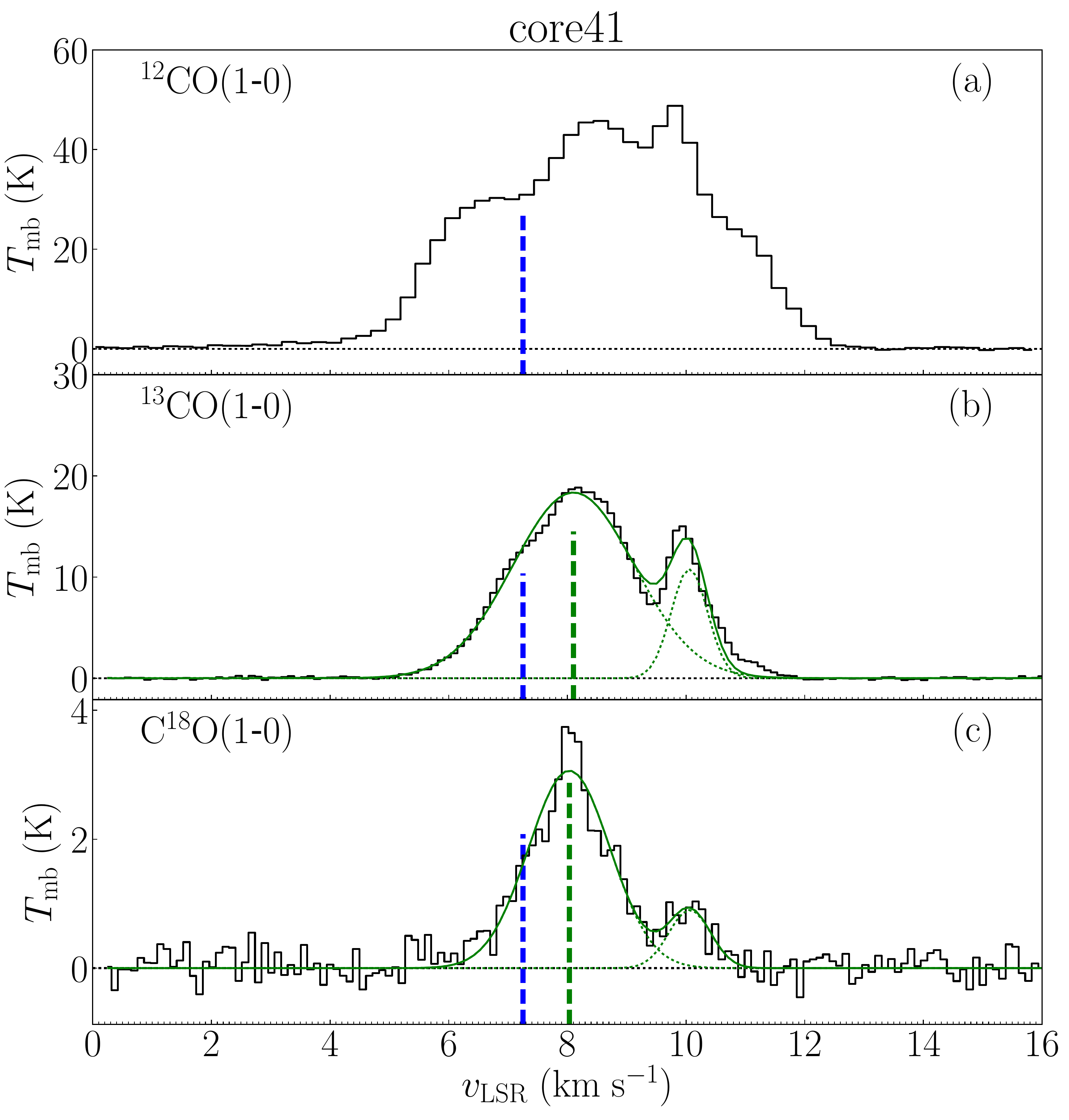}
\caption{
Same as Figure \ref{fig:core5}, but for core 41.
The green dashed line shows the centroid 
velocity $v_{\rm gauss}$ of the Gaussian component
that is closest to $v_{\rm NH_3}$.
The green dotted curves show the fitting to
each Gaussian component.
\label{fig:core41}}
\end{figure}

Figure \ref{fig:core5} shows the spectra of 
core number 5 in the Kirk17 sample.
We show the spectra of $^{12}$CO(1-0), 
$^{13}$CO(1-0), and C$^{18}$O(1-0) in black.
In order to perform the Gaussian fit, 
we visually identify the number of Gaussian
components for each spectrum and provide
the initial guess for amplitude,
centroid velocity, and velocity dispersion for each spectrum.
Then, we use the {\tt scipy.optimize.curve\_fit} 
function from the python package Scipy
to perform single-/multi-Gaussian fitting
to $^{13}$CO and C$^{18}$O spectra.
If there is only one component,
we take the centroid velocity $v_{\rm gauss}$
as the envelope velocity. If more than one velocity component
is identified, we take the one that is nearest to 
$v_{\rm NH_3}$ as the  envelope velocity\footnote{This
could introduce a bias toward a small difference
between core and envelope velocity.
However, we argue that this is unlikely.
Assuming the intensity ratio NH$_3$/$^{13}$CO is proportional to
the abundance of NH$_3$, a large difference between 
the velocity centroids of NH$_3$ and $^{13}$CO would
result in relatively high NH$_3$ abundance 
since the peak of the NH$_3$
component would be coincident with the $^{13}$CO line wing.
Moreover, we select cores with
only one CO velocity component and derive the core-to-envelope velocity dispersion, $\sigma_{\rm ce}$, 
and obtain $\sigma_{\rm ce}$($^{13}$CO) = 0.28 km s$^{-1}$ and
$\sigma_{\rm ce}$(C$^{18}$O) = 0.17 km s$^{-1}$. 
This is very similar to the values obtained using the entire sample of cores,
hence we believe that including multi-component spectra does not 
produce any bias in our results.}, similar to the
study by \citet[][see their discussions in section 5
and figure 6]{2007ApJ...668.1042K}.

Core 5 (Figure \ref{fig:core5}) shows one Gaussian
component in $^{13}$CO and C$^{18}$O spectra.
However, a significant fraction of the cores
show complex velocity features.
In Figure \ref{fig:core41}, we show another example,
core 41. This core has roughly three velocity 
components in the $^{12}$CO(1-0) line and two components
in $^{13}$CO(1-0) and C$^{18}$O(1-0). 
$v_{\rm NH_3}$ of this core does not corresponds to 
any of the carbon monoxide line components.
Based on our visual identification, 100 out of
the 237 cores from Kirk17
show multiple velocity components in either the
$^{13}$CO or the C$^{18}$O spectrum (or both).
$^{12}$CO spectra typically show more complex profiles. 

\begin{figure*}[htbp]
\epsscale{0.55}
\plotone{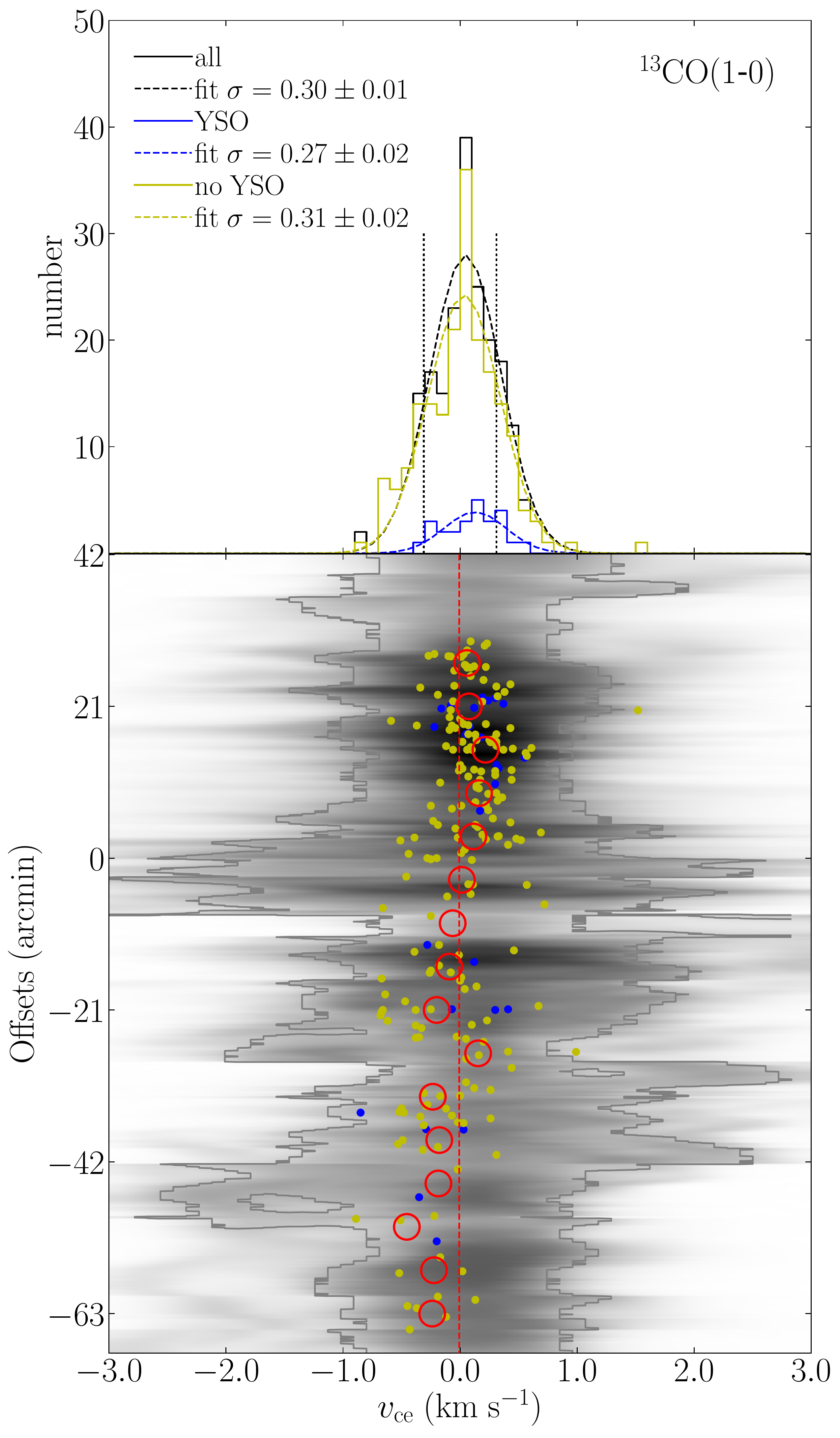}
\plotone{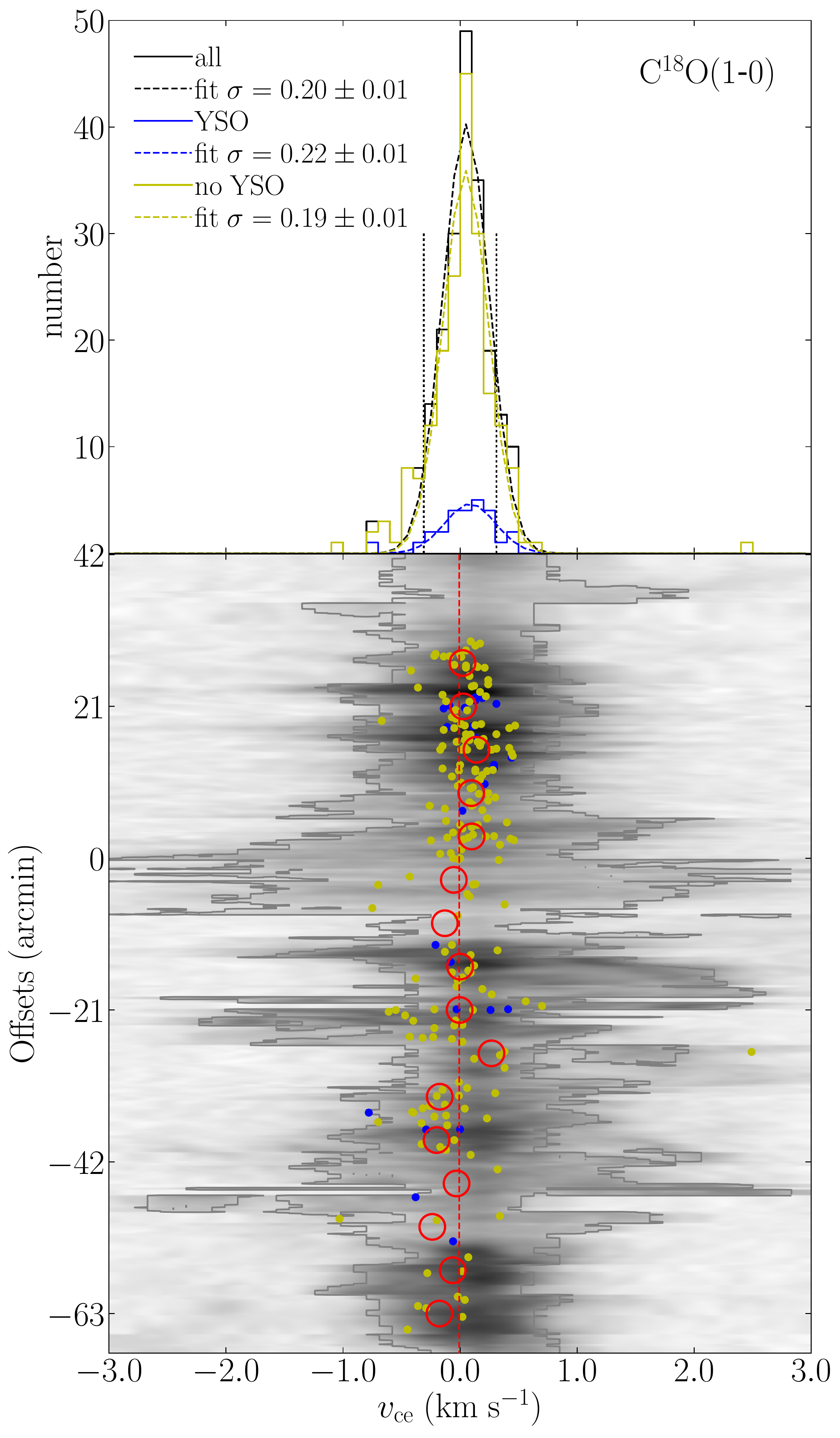}
\caption{
{\bf Left:} 
{\it Top panel:} Histograms of core-to-envelope
velocity differences $v_{\rm ce}$($^{13}$CO).
The black histogram shows the distribution
for the entire Kirk17 core sample. The black dashed 
curve shows the Gaussian fit to the black histogram.
The blue histogram and dashed curve correspond to 
the cores with YSO; the yellow without YSO.
The dispersion values obtained from Gaussian fits to the 
histograms are shown in the upper left corner.
The vertical dotted lines show the sound speed.
{\it Bottom panel}: 
The gray background shows the PV-diagram along the 
orange curve in Figure \ref{fig:filregions}.
The zero offset is the big red dot on the orange curve
between OMC-1 and ONC. Positive offset is toward
the north (OMC-2/3 direction). Note that
21 arcmin corresponds to $\sim$2.5 pc at 400 pc.
At each offset, the PV-diagram zero velocity is
defined at the intensity peak. The gray contour
denotes the half maximum at each offset. The yellow
dots denote the starless cores and the blue dots
denote the cores with YSO. The offset (y-axis) is along
the ISF spine (orange curve). The empty red circles
are offset bins every 6\arcmin. Their $v_{\rm ce}$
values are averaged within the bins.
{\bf Right:}
Same as left, but for $v_{\rm ce}$(C$^{18}$O).
\label{fig:vdiffgauss}}
\end{figure*}


To estimate possible systematic difference
between the dense core velocity $v_{\rm NH_3}$ 
and the envelope gas velocity, we calculate for each core
the 1D core-to-envelope velocity 
$v_{\rm ce} \equiv v_{\rm NH_3}$ - $v_{\rm gauss}$
using the $^{13}$CO(1-0) and C$^{18}$O(1-0)
line Gaussian fitting results.
In the top panels of Figure \ref{fig:vdiffgauss},
we show the distributions of $v_{\rm ce}$($^{13}$CO)
and $v_{\rm ce}$(C$^{18}$O) (black bins), respectively.
The bin size is 0.1 km s$^{-1}$. We then fit a
Gaussian to the $v_{\rm ce}$ distribution. 
We can see the 1D core-to-envelope
Gaussian dispersion $\sigma_{\rm ce}$ is
0.20$\pm$0.01 km s$^{-1}$ for C$^{18}$O and
0.30$\pm$0.01 km s$^{-1}$ for $^{13}$CO.
We also apply the maximum likelihood method
(using {\tt scipy.stats.norm.fit})
to estimate $\sigma_{\rm ce}$, which gives
$\sigma_{\rm ce}$($^{13}$CO) = 0.33 km s$^{-1}$ and
$\sigma_{\rm ce}$(C$^{18}$O) = 0.25 km s$^{-1}$ 
(removing the outlier beyond 2 km s$^{-1}$; 
if not, $\sigma_{\rm ce}$ = 0.30 km s$^{-1}$).


We investigate the effect of varying envelope sizes
by extracting envelope spectra from a number of circular
regions centered on the core but with different radii.
For this purpose we use the original 
CARMA-NRO Orion Survey data cubes
(i.e., before smoothing to 32\arcsec\ resolution).
At each core location, we extract $^{13}$CO and C$^{18}$O
spectra from circular regions with diameters 8\arcsec,
16\arcsec, 32\arcsec, 64\arcsec, and 128\arcsec, i.e.,
factors of 2 increase from the beam size 
(8\arcsec\ corresponds to a physical scale of 0.015 pc).
The respective fitting results of the 
core-to-envelope dispersion for $^{13}$CO are
0.29 km s$^{-1}$, 0.30 km s$^{-1}$, 0.30 km s$^{-1}$, 
0.34 km s$^{-1}$, 0.39 km s$^{-1}$; for C$^{18}$O they are
0.26 km s$^{-1}$, 0.25 km s$^{-1}$, 0.25 km s$^{-1}$, 
0.26 km s$^{-1}$, 0.32 km s$^{-1}$. As we compare 
the core velocity to that of the envelopes averaged over a larger area,
the velocity dispersion becomes slightly higher.

\begin{figure}[htbp]
\epsscale{1.}
\plotone{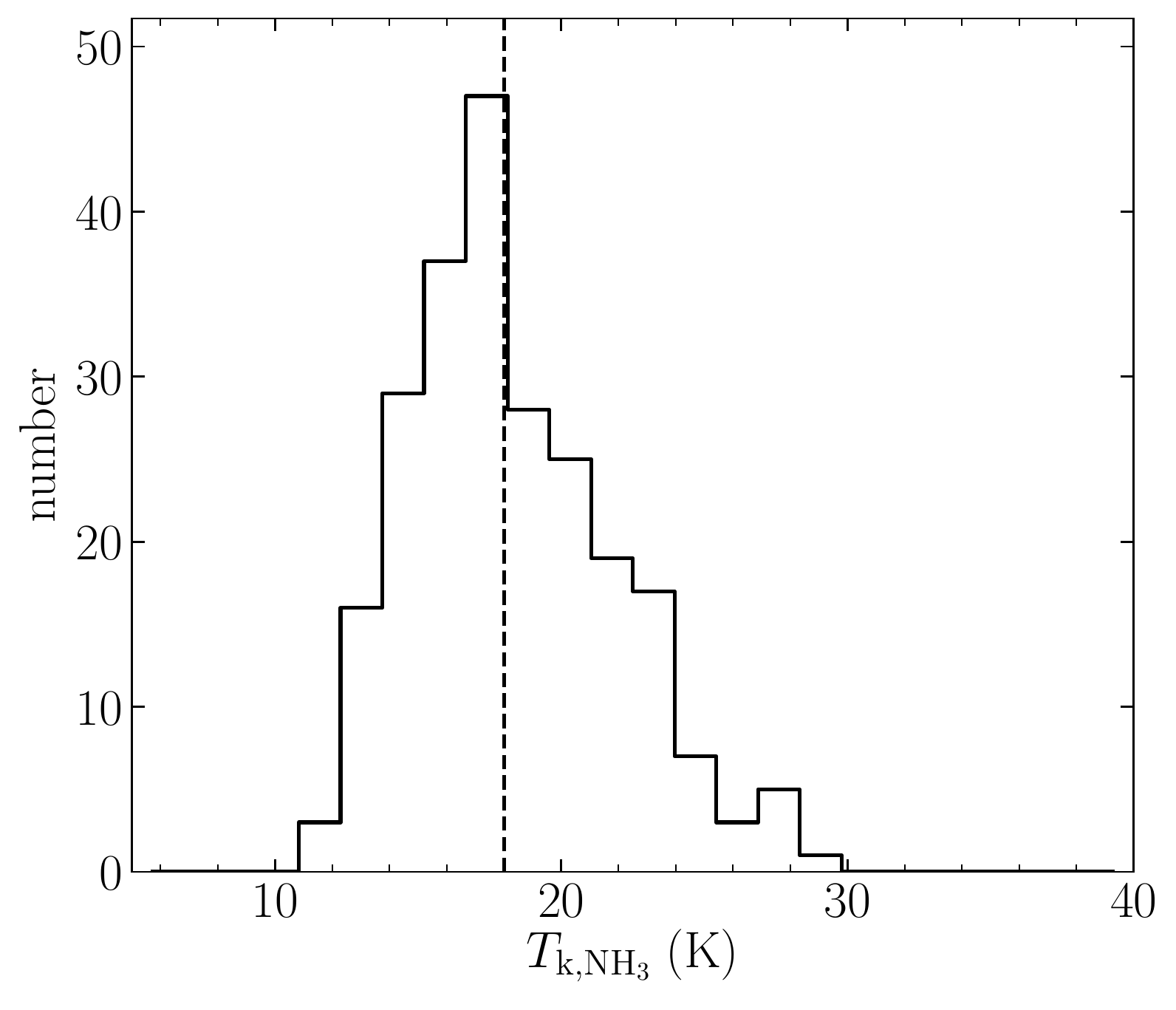}
\caption{
Histogram of core kinetic temperature derived from
NH$_3$ inversion lines. The data are taken from 
Kirk17. The vertical dashed line shows the median 
temperature (18 K) used to compute the sound speed.
\label{fig:tkinhist}}
\end{figure}

\begin{figure}[htbp]
\epsscale{1.1}
\plotone{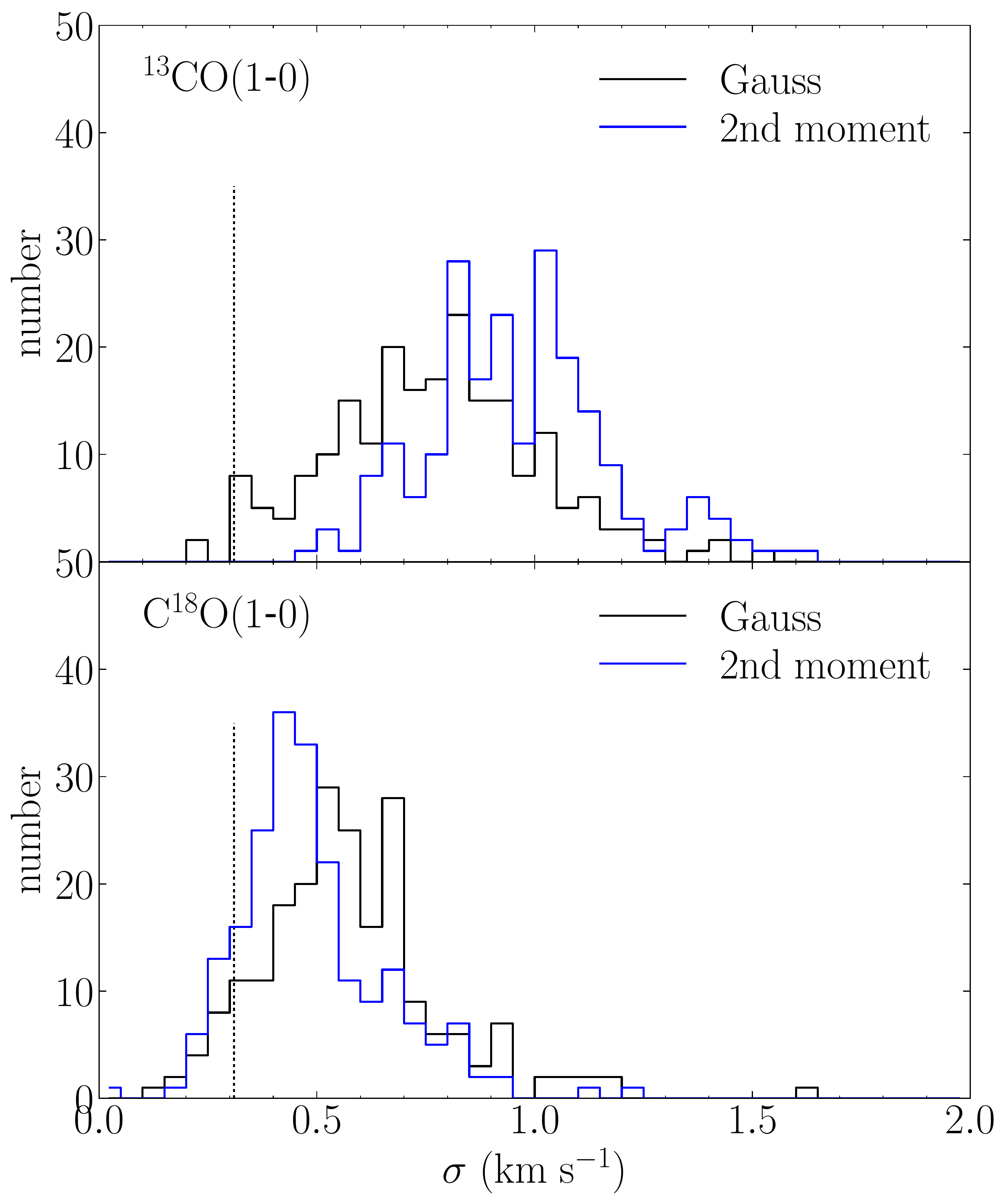}
\caption{
Histograms showing the distribution of the line dispersion in the $^{13}$CO (top panel) and C$^{18}$O (bottom panel) spectra from cores. 
The black histograms show the distribution of   dispersion values obtained from gaussian fits to the spectra, while the blue histograms show the distribution of values obtained from 
the second moment of the spectra. The vertical dotted line shows
the sound speed calculated at 18 K. 
\label{fig:sigmahist}}
\end{figure}

In the top panels of Figure \ref{fig:vdiffgauss}
 we indicate the sound speed,
calculated using a gas temperature of 18 K
\citep[following appendix A1 in][]{2017A&A...599A..99O}, 
by the vertical dotted lines. Figure \ref{fig:tkinhist}
shows the distribution of core gas kinetic temperatures
derived from NH$_3$ data (Kirk17).
The mean temperatures are 18.3 and 17.8 K, respectively, while the 
minimum and maximum temperature are 11.3 and 29.3 K,
respectively We adopt 18 K as
the representative temperature for the sound speed 
calculation (which results in a value of 0.31 km s$^{-1}$). 
For C$^{18}$O, the sound speed is at 1.55$\sigma_{\rm ce}$,
which from the Gaussian fitting to $v_{\rm ce}$(C$^{18}$O)
would imply that  about 88\% of the cores have relative motion
slower than the local sound speed (note that from the data,
the actual fraction of  cores
with $v_{\rm ce}$(C$^{18}$O) $<$ 0.31 km s$^{-1}$ is 78\%.)
For $^{13}$CO, the sound speed is at 1.0$\sigma_{\rm ce}$.
Based on the Gaussian fitting, this implies that 68\% of 
the cores have relative motion slower than the local
sound speed (the actual fraction of cores
with $v_{\rm ce}$($^{13}$CO) $<$ 0.31 km s$^{-1}$ is 66\%).
Note that the velocity dispersion and sound speed
discussed here are both 1D quantities.

Figure \ref{fig:sigmahist} shows the distribution of
line velocity dispersions for $^{13}$CO(1-0) and C$^{18}$O(1-0)
(black bins). Again, we indicate the sound speed
as the vertical dotted line. The figure shows that the majority 
of cores are located in regions where the carbon monoxide
gas line dispersion is supersonic. Therefore,
the core-to-envelope velocity dispersion $\sigma_{\rm ce}$
is significantly smaller than the local line width.
In the bottom panels of Figure \ref{fig:vdiffgauss},
we show the core-to-envelope velocities on top of
the PV-diagram along the ISF ridge defined in K18
and Figure \ref{fig:filregions}. 
As can be seen, for the vast majority of cores,
the core-to-envelope velocities 
$v_{\rm ce}$($^{13}$CO) and $v_{\rm ce}$(C$^{18}$O) are 
predominantly smaller than the velocity FWHM at 
the position of the core (shown as gray contours in 
Figure \ref{fig:vdiffgauss}),
confirming the results from the Gaussian fittings.
Through a study of the kinematics in the Orion Nebula Cluster (ONC), 
\citet{2009ApJ...697.1103T} showed that the stars and
$^{13}$CO gas are kinematically tied, consistent with
our results for dense cores.

We made separate histograms of $\sigma_{\rm ce}$ for cores
with and without YSOs in order to investigate 
whether there is a difference between
pre- and protostellar cores as a result of evolution.
The YSO-core assignment is given by Lane16,
where they checked against catalogs from 
\citet[][Spitzer observation]{2012AJ....144..192M} 
and \citet[][Herschel observation]{2013ApJ...767...36S}.
In Figure \ref{fig:vdiffgauss}, we show $v_{\rm ce}$
distribution for protostellar cores as blue histograms;
starless cores as yellow histograms. 
The $v_{\rm ce}$ dispersions between the two populations
are different by 10\%. The $v_{\rm ce}$ centroid
for the starless cores is 0.03$\pm$0.02 km s$^{-1}$
in the $^{13}$CO fitting; for the protostellar cores 
0.13$\pm$0.02 km s$^{-1}$. In the C$^{18}$O fitting,
the $v_{\rm ce}$ centroid
for the starless cores is 0.04$\pm$0.01 km s$^{-1}$; 
for the protostellar cores 0.08$\pm$0.01 km s$^{-1}$. The cores
with YSOs tend to have an excess of positive core-to-envelope
velocities. However, note the limited number of cores with 
associated YSOs (see the blue dots in Figure 
\ref{fig:vdiffgauss}).

\subsection{Filament and core dynamics}\label{subsec:core}

\begin{figure*}[htbp]
\epsscale{1.15}
\plotone{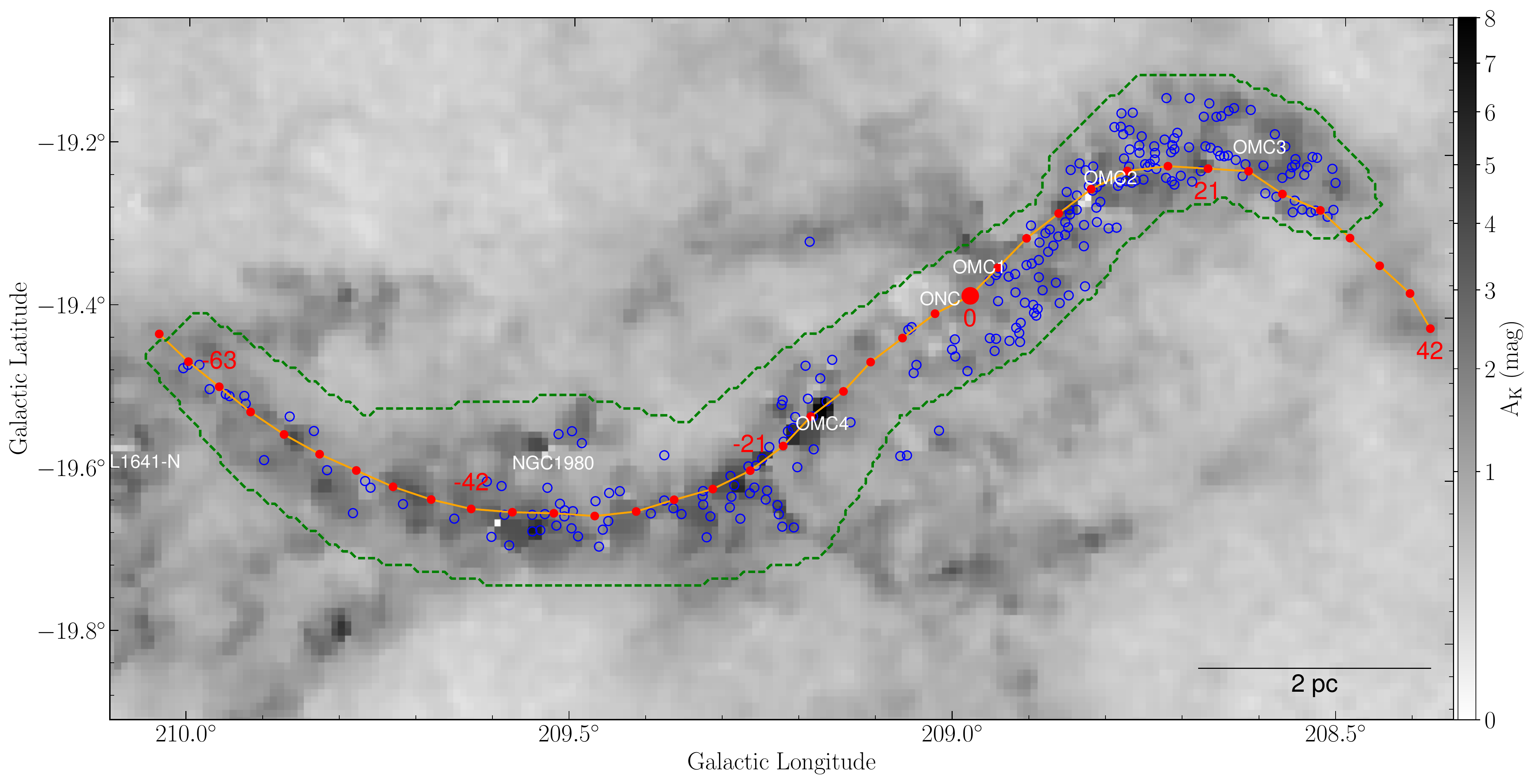}
\caption{
Cores on the Integral-Shaped Filament in Orion A.
The gray-scale background image shows the 
extinction map from \citet{2018A&A...614A..65M}.
The blue circles mark the position of the Kirk17 cores.
The green dashed contour shows the filament
defined for estimating $m_l$ in \S\ref{subsec:core}.
The orange curve defines the PV cut following figure 5
of \citet{2018ApJS..236...25K} but extending further
to cover the entire core sample in the east.
The red dots along the curve indicate 3\arcmin\ segments.
The red numbers show the offsets in Figure \ref{fig:pvcores}.
\label{fig:filregions}}
\end{figure*}

We re-analyze the filament virial status in
the context of the model by \citet{2000MNRAS.311...85F}. 
In this model, the virial velocity for an unmagnetized 
filamentary cloud is related to the mass per unit 
length $m_l$ by: 
\begin{equation}\label{eq:sigmavirial}
\sigma_{\rm vir} = \sqrt[]{\frac{Gm_l}{2}},
\end{equation}
where G is the gravitational constant.

\citet{2000MNRAS.311...85F} used the Orion A 
$^{13}$CO data with a resolution of 1.7\arcmin\
from \citet{1987ApJ...312L..45B}
to derive a value of $m_l
= 355$ M$_\odot$ pc$^{-1}$  for the ISF. 
We use the extinction map \citep{2018A&A...614A..65M} 
to estimate the ISF mass. Figure \ref{fig:filregions}
shows our definition of the filamentary region.
We define the entire ISF filament region 
to roughly cover the Kirk17 core sample
(green dashed polygon). The total mass of the filament is 
$\sim$3835 M$_\odot$ and the total length is $\sim$12 pc.
The length is derived from a visually-defined curve that 
follows the ridge line of the ISF 
(originally defined in figure 5 of K18, but extended
to the eastern end of the filament). 
The recent GAIA data show that the ISF 
is likely parallel to the plane of the sky 
\citep{2018A&A...619A.106G}, although with 
a few pc variation \citep{2018arXiv180711496S}.
Hence, we do not correct the length of
the filament due to inclination effects.
The ISF $m_l$ we derive is 320 M$_\odot$ pc$^{-1}$,
a factor of 0.90 smaller than \citet{2000MNRAS.311...85F}.
This falls in the $m_l$ range (125-800 M$_\odot$ pc$^{-1}$) 
estimated by \citet[][see their figure 5]{2016A&A...590A...2S}.
Our estimated virial velocity for an unmagnetized filament is 
$\sigma_{\rm vir}\sim$0.83 km s$^{-1}$. 

\begin{figure*}[htbp]
\epsscale{0.8}
\plotone{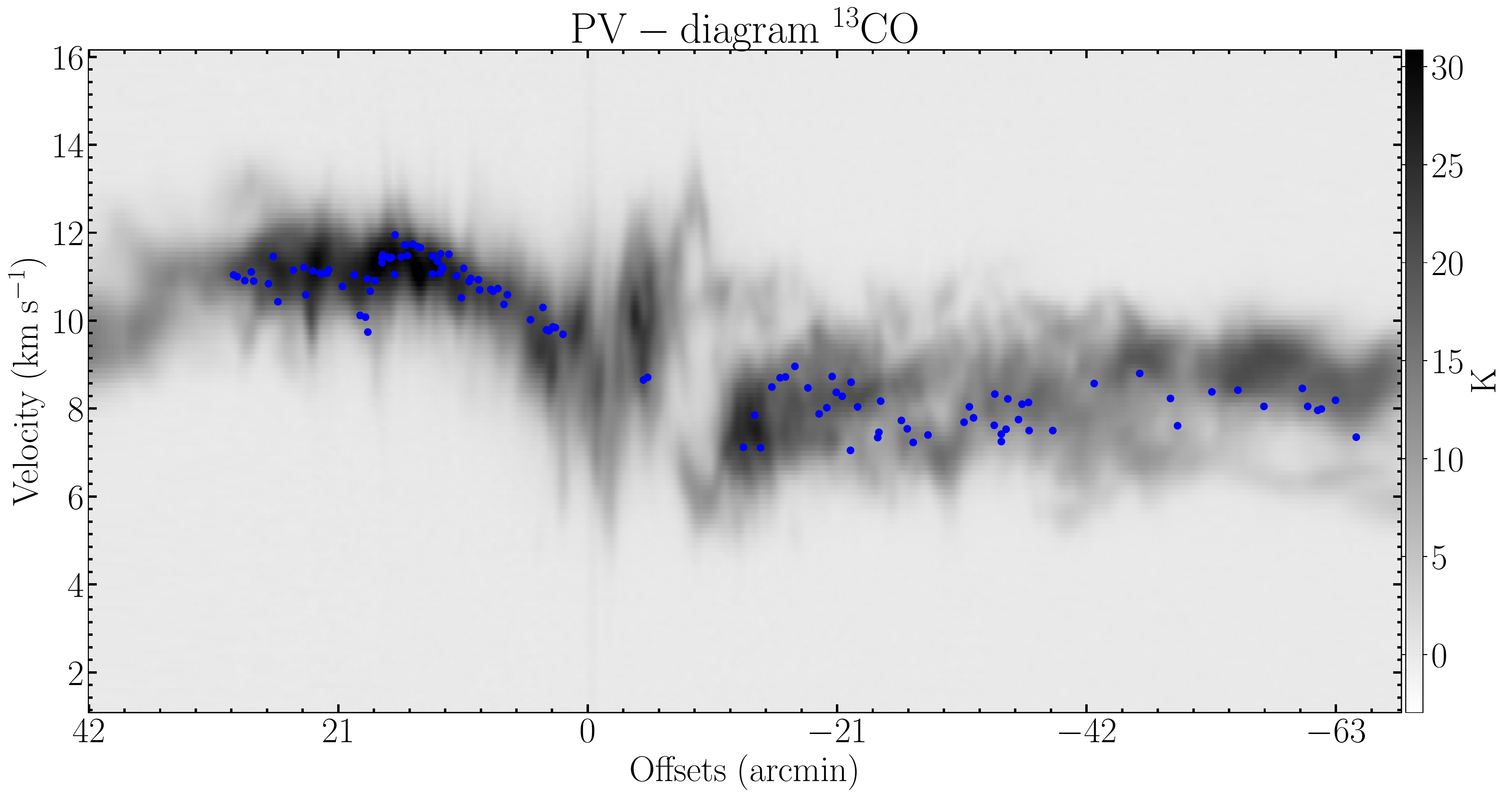}\\
\plotone{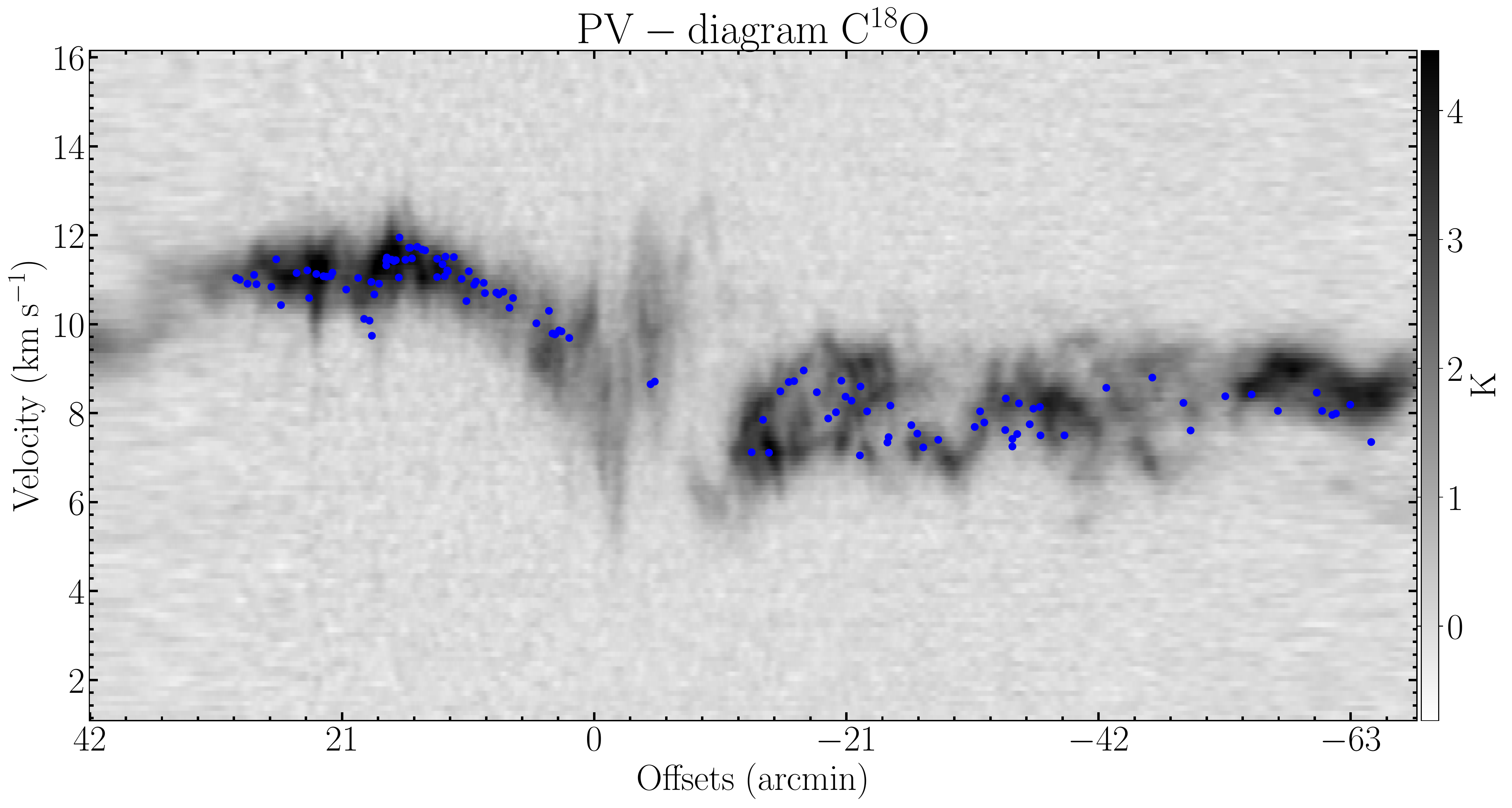}
\caption{
Position-velocity (PV) diagrams along the ISF using the $^{13}$CO (top)
and C$^{18}$O (bottom) data. The PV cut is defined
as the orange curve along the ISF spine with a width of 
3\arcmin (Figure \ref{fig:filregions}). The zero-offset
position is the large red dot between OMC1 and ONC
in Figure \ref{fig:filregions}. The positive
offset is toward the west direction (OMC-2/3).
Note that 21 arcmin corresponds to $\sim$2.5 pc at 400 pc.
The blue dots show the Kirk17 cores.
The velocity is different from Figure \ref{fig:vdiffgauss}
bottom panels where the zero velocity is the gas intensity peak.
\label{fig:pvcores}}
\end{figure*}

Recall from \S\ref{subsec:kin} that the core-to-envelope
velocity dispersion $\sigma_{\rm ce}$ for $^{13}$CO and C$^{18}$O is 0.30 and 0.20 km s$^{-1}$, respectively. 
Applying a factor of $\sqrt[]{3}$, the 3D dispersion for
$^{13}$CO is 0.52 km s$^{-1}$ and for C$^{18}$O is 0.35 km s$^{-1}$.
Both numbers are smaller than the $\sigma_{\rm vir}$ in
the unmagnetized scenario, indicating that the cores 
are bound to the ISF filament. 
In the presence of magnetic fields, $\sigma_{\rm vir}$ 
depends on both the poloidal field in the filament
and the helical field that wraps the filament. More
robust measurements of the field strength are needed
\citep[note the recent progress
by][]{2017ApJ...846..122P,2018A&A...614A.100T}.

Note that the ISF has an overall velocity gradient and
the core kinematics follow the trend.
To illustrate this, we make PV-diagrams along the
ISF for $^{13}$CO and C$^{18}$O, shown in Figure \ref{fig:pvcores}.
The PV-cut follows the orange curve from Figure \ref{fig:filregions}.
The offsets in Figure \ref{fig:pvcores} are labeled
as the red numbers in Figure \ref{fig:filregions}.
We also overplot the cores on the PV-diagram. As one can see,
the ISF shows a wave-like velocity structure
\citep[see K18 figures 20-22; also see][]{2016A&A...590A...2S}
and the cores closely follow the ISF kinematics. 
The overall velocity distribution of the ISF results in
a super-virial core-to-core velocity dispersion of 2.92 
km s$^{-1}$, which is different from what was seen in Perseus
\citep[sub-virial core-to-core
dispersion,][]{2010ApJ...723..457K,2015ApJ...799..136F}.
While the final fraction of gas in Orion A and Perseus
that is deposited into stars (SF efficiency) is unclear, 
the forthcoming star cluster in Orion A is less likely to
be bound compared to the cluster in Perseus. 
This emphasizes the importance of the initial condition
for star cluster formation.


\section{Discussion}\label{sec:discussions}

\subsection{Comparing core formation capability between Orion A and IRDC G28}\label{subsec:cf}

Within the range 0.08 $\la\Sigma_{\rm ex}\la$ 0.28 g cm$^{-2}$,
Orion A appears to be more capable of forming
dense cores than G28 (Figure \ref{fig:dpfcompare}).
In \S\ref{subsec:dpfirdc} we suggested that this could be caused by SF
feedback in Orion A, although we did not indicate the main 
feedback mechanism that could be responsible for this.
Outflows are unlikely the source for the
difference between Orion A and G28, as both clouds have widespread outflows
\citep{2009A&A...496..153D,2017ApJ...837...60B,2019ApJ...874..104K}.
The main difference likely comes from the presence of
massive star feedback in Orion A (the region of G28 studied 
in \citet{2019ApJ...874..104K} is not impacted by massive stars).
In Figure \ref{fig:tkinmap} we have shown that
the dense gas in Orion A is not significantly heated,
thus the heating from ionizing photons is not likely 
dominating the difference between these two clouds.
We suggest that pressure from the expanding HII region
\citep{2019Natur.565..618P}, which possibly compresses
the cloud and generates more dense gas, 
may be responsible for the higher incidence of dense
cores in Orion A. \citet{2018ApJ...862..121F} reported
expanding shells and bubbles in Orion A, which may also
be important in generating dense gas.
Our results emphasize the important 
role of stellar feedback in SF.

\subsection{ISF core group dynamics}\label{subsec:coredyn}

The sub-sonic core-to-envelope kinematic feature we detect in Orion A 
is seen in other nearby star-forming clouds 
\citep[i.e., Taurus, Perseus, Ophiuchus, see][]{2004ApJ...614..194W,
2007ApJ...668.1042K,2010ApJ...723..457K,2015ApJ...799..136F},
although the Orion A cloud is quite different from these clouds in terms of
active massive SF and destructive feedback. 
These suggest that dense
molecular cores remain tied to their host clouds
during their formation regardless of the environment. 

The kinematics of cores can be useful in testing SF theories.
For instance, in the competitive accretion model
\citep{2001MNRAS.323..785B,2009MNRAS.400.1775S},
protostars and the host cloud globally collapse toward the
cloud's gravitational potential minimum. 
As shown by \citet{2001MNRAS.323..785B},
protostars compete for the gas through tidal accretion
and the relative velocity between a protostar and the surrounding gas 
is sub-sonic throughout the global collapse.
However, it is to be clarified how the cores 
seen in observations are linked to
the structures seen in the simulations. 
For instance, detailed radiative transfer modeling
may be necessary to show if the tidal sphere in simulations
corresponds to the dense cores in observation.
On the other hand,
the fact that dense cores have low core-to-envelope 
velocity dispersion is a fundamental prediction of 
the ``colliding flow'' model in \citet{2001ApJ...553..227P}.
Cores do not move freely in the envelope because
they form in the post-shock gas
at the intersection of colliding flows
(i.e., in regions where the flow velocity 
stagnates, as it is dissipated in shocks).

\subsection{The dynamical status of the ISF}

The work by \citet{2009ApJ...697.1103T} have shown the 
similarity in the kinematics between the stars 
and $^{13}$CO gas in the ISF.
Subsequently, \citet{2016A&A...589A..80H,2016A&A...590A...2S,
2019MNRAS.tmp.1395G}  explored the kinematic link
between YSOs and the cloud gas in Orion A. 
Together with our findings, these results suggest that 
SF is ongoing in the ISF and that the low 
velocity dispersion in the protostars 
relative to their local gas probably originates from 
the low core-to-envelope velocity reported here. 
However, the explanation of the origin and the future 
of the ISF remains unclear, and various hypotheses
have been proposed.

\citet{2009ApJ...697.1103T} argued that the ISF
cloud and stars are undergoing a global collapse
toward the ONC. The basis for their
argument was the observed velocity gradient along the
declination direction.  The gas kinematics 
and the radial velocity of the young stars 
show a change toward red-shifted velocities 
in the OMC-2/3 regions north of the ONC
(their figure 3), while the southern ISF shows 
an approximately flat velocity distribution. 
\citet{2009ApJ...697.1103T} thus argued that the ISF is
collapsing toward the ONC, with the OMC-2/3 regions 
located on the near side but having red-shifted velocities. 

With a higher resolution (30\arcsec) N$_2$H$^+$ data,
\citet{2017A&A...602L...2H} revisited the topic.
They also argued that the ISF is undergoing 
gravitational collapse toward the ONC. 
However, their picture supports 
a collapse from the far side. The basis for their argument 
is the blue-shifted velocity gradients in both the north
and south of the cluster (see their Figure 1c) combined with the
inference that the ONC must be on the near side of the ISF.  
Note that the collapse scale in \citet{2017A&A...602L...2H}
is smaller than that discussed in \citet{2009ApJ...697.1103T}.

Recent analyses with APOGEE and Gaia surveys have
shown that the ONC is slowly expanding
\citep{2018AJ....156...84K,2019ApJ...870...32K}, raising
the question that whether gas is falling into ONC.
However, it is possible if the ISF collapsing direction 
is toward OMC-1 (the densest gas region in the ISF),
as \citet{2019MNRAS.tmp.1395G} recently 
have suggested that the stars in the OMC-1 region
are undergoing gravitational contraction.
On the other hand, \citet{2019MNRAS.tmp.1395G}
showed that OMC-4/5 are not moving toward OMC-1.
This raises the question whether the ONC/OMC-1
region is the gravitational focus.
Meanwhile, it is possible that the global collapse
is strongly impacted by the intense feedback in the area,
resulting in a very complex environment.

Alternatively, \citet{2016A&A...590A...2S} and
\citet{2018MNRAS.473.4890S} proposed the 
``slingshot mechanism'' for the formation of stars 
in ISF of Orion A.  They emphasized the wave-like
geometry of the ISF and postulated that the filament
has an oscillatory motion. In this scenario, 
cores move with the filament and the newly-formed
stars are ejected from the oscillating filament once they are 
massive enough to decouple from the dense gas. 
The model addressed the wave-like
spatial and kinematic morphology of the ISF and 
provided another explanation of the high velocity 
of newly-formed stars. Recently, \citet{2018arXiv180711496S}
used Gaia data to suggest that the ISF filament is a 
standing wave.

With our newly acquired CARMA-NRO Orion gas data,
but making the PV-diagram along the ISF (instead of 
declination, see Figure \ref{fig:filregions}), 
we detect more complicated kinematic features
(Figure \ref{fig:pvcores}). The cloud shows a ``wave-like''
feature (more so in the north with positive offsets in Figure
\ref{fig:pvcores} than in the south). Such kinematics 
present a challenge in terms of simply explaining
the ISF with a global collapse toward ONC. Moreover,
the northern part shows an oscillatory feature at
scales of $\sim$ 3\arcmin. All these results suggest
rather complex cloud kinematics and more information
is needed to understand the origin and evolution of the ISF.

We notice something interesting from the core-to-envelope
PV-diagrams in Figure \ref{fig:vdiffgauss}. In the bottom panels,
we bin the core-to-envelope velocities along the ISF.
Each empty red circle represents the average core-to-envelope
velocity within a 6\arcmin\ bin. Along the ISF,
the distribution of the red circles is again oscillatory. 
Note that this is 
not the wave-like structure of the filament gas.  
As shown in the figures, the cores are red-shifted 
relative to the gas in the OMC-2/3 regions
(positive offsets), while in the southern
region (negative offsets) the cores are blue-shifted 
relative to the gas. The origin of the core motion is
unclear, and this is the first time they are
detected. It is not clear why one should expect
an oscillatory motion in velocities between
the core and the gas, but it does remind us of
the slingshot mechanism because of the ``wave''.
The expanding bubble from the Trapezium cluster may
push away low-density envelopes \citep{2019Natur.565..618P},
resulting in an apparent shift of the core-to-envelope
velocity. However, this scenario would require the 
impact of the feedback at very high extinctions and
far away from the location of massive stars along the entire 
ISF \citep[see discussions in][]{2018A&A...619A.106G},
as the cores are deeply embedded.  

\section{Conclusions}\label{sec:conclusions}

In this paper, we studied sub-mm continuum core formation
and kinematics in the Orion A molecular cloud. Various
sources of publicly available and proprietary data were combined with
our CARMA-NRO Orion Survey data for the study.
Below we list our major conclusions.

\begin{enumerate}

\item We have compared 850 $\mu$m continuum core detections 
\citep{Lane2016} with NIR extinction \citep{2018A&A...614A..65M}.
We find a detection threshold of $A_V\approx$ 5-10 mag
for the cores within the declination range of
(-9.5\arcdeg, -4.5\arcdeg). Together with previous findings 
in Ophiuchus and Perseus, we suggest a universal core
formation extinction threshold of $A_V\sim$ 5-10 mag for clouds
within 500 pc from the Sun. This threshold also appears to be
consistent with the star formation extinction threshold 
suggested by \citet{2008ApJ...680..428G,2010ApJ...724..687L,
2010ApJ...723.1019H}. The total 850 $\mu$m core mass is 
about 4\% of the mass of the entire Orion A cloud.

\item A pixel-by-pixel comparison between the sub-mm
continuum and the NIR extinction shows a steady increase
in the continuum detection probability for $A_V$
between $\sim$ 5 and 20 mag, and a potential turning point
at $A_V\sim$20 mag where the slope in the relation decreases.
This turning point is also seen in continuum core detection.

\item We have compared the pixel-based
continuum detection in Orion A with the infrared dark cloud G28.37+0.07.
Within the cloud extinction range of 15 mag $\la A_V \la$ 60 mag,
Orion A shows more dust continuum emission detection,
suggesting Orion A has been able to form more dense gas
($\Sigma_{\rm mm}\ga$ 0.044 g cm$^{-2}$). 
Our study shows that the feedback from high-mass stars in 
Orion A, possibly expanding HII regions, is a potential 
cause of the difference between these two clouds.
The continuum detection probability function (DPF) 
can be useful in showing the differences
in core formation between clouds. 

\item We have investigated the dense core kinematics within
the Orion A integral-shaped filament (ISF). 
In particular, we compare
the velocities of the dense cores (traced by NH$_3$) 
with the velocity of their 
surrounding lower density envelope 
(traced by $^{13}$CO and C$^{18}$O). 
We find that the 1D dispersions in the distribution of 
the core-to-envelope velocity are 0.20 and 0.30 km s$^{-1}$, 
when using C$^{18}$O and
$^{13}$CO to trace the envelope, respectively.
These are comparable to the sound speed of the dense gas
and smaller than the local line dispersion in
the $^{13}$CO(1-0) and C$^{18}$O(1-0) spectra.
This is consistent with previous findings
in Perseus, suggesting that the dense molecular
cores are kinematically tied to their parent cloud, 
independent of the cloud environment.
However, compared to Perseus, the ISF core-to-core
velocity dispersion is super-virial, largely due to
the overall velocity gradient in the ISF.
This difference may impact the dynamics of 
the forthcoming cluster in the future, which emphasizes
the importance of initial cloud conditions for 
star cluster formation.

\item The cloud gas velocity and the core-to-envelope 
velocity both show wave-like features in PV-diagrams,
although the two are conceptually different. 
Their relation to the overall dynamical picture of 
the ISF is to be determined, 
but see \citet{2018MNRAS.473.4890S}
for recent work addressing the wave-like signature in the ISF. 
The kinematic difference between the dense cores and the
cloud presented in this work remains to be theoretically addressed. 

\end{enumerate}

\software{The data analysis in this paper uses python packages Astropy \citep{Astropy-Collaboration13}, SciPy \citep{scipy}, and Numpy \citep{numpy}. The figures are plotted using python pacakges APLpy \citep{Robitaille12} and Matplotlib \citep{matplotlib}.}

\acknowledgments 
We thank the anonymous referee for constructive reports that
significantly improve the quality of the paper.
We acknowledge the fruitful discussions with James Lane and Helen Kirk.
SK and HGA were funded by NSF award AST-1140063 while conducting this study.
RSK thanks for support from the German Science Foundation (DFG) via the Collaborative Research Center SFB 881 ``The Milky Way System'' (subproject B1 and B2).
AS acknowledges funding through Fondecyt regular (project code
1180350), ``Concurso Proyectos Internacionales de Investigaci\'on''
(project code PII20150171), and Chilean Centro de Excelencia en
Astrof\'isica y Tecnolog\'ias Afines (CATA) BASAL grant AFB-170002.
P.P. acknowledges support by the Spanish MINECO under project AYA2017-88754-P. R.J.S. gratefully acknowledges an STFC Ernest Rutherford fellowship (grant ST/N00485X/1).
The JCMT has historically been operated by the Joint Astronomy Centre on behalf of the Science and Technology Facilities Council of the United Kingdom, the National Research Council of Canada and the Netherlands Organisation for Scientific Research. Additional funds for the construction of SCUBA-2 were provided by the Canada Foundation for Innovation. The identification number for the programme under which the SCUBA-2 data used in this paper is MJLSG313.
This research used the facilities of the Canadian Astronomy Data Centre operated by the National Research Council of Canada with the support of the Canadian Space Agency. 
This research was carried out in part at the Jet Propulsion Laboratory which is operated for NASA by the California Institute of Technology.

\facilities{CARMA, No:45m, JCMT, GBT, VISTA, Herschel}

\bibliographystyle{aasjournal}
\bibliography{ms}

\end{document}